\documentclass{article} 
\usepackage{arxiv,times}

\usepackage{hyperref}
\usepackage{url}
\usepackage{amsmath}
\usepackage{xcolor}
\newcommand{\added}[1]{#1}

\usepackage{amsmath,amssymb}
\usepackage{bm}
\usepackage{graphicx}
\usepackage{adjustbox}
\usepackage{tabularx}
\usepackage{multirow}
\usepackage{rotating}
\usepackage{placeins}
\usepackage{subcaption}
\usepackage{natbib}
\usepackage{microtype}
\usepackage{graphicx}
\usepackage{caption}
\usepackage{subcaption}
\usepackage{booktabs} 
\usepackage{breqn}
\usepackage{amsmath}
\usepackage{bm}
\usepackage{tikz}
\usepackage{pifont}

\newcommand{\cmark}{\ding{51}}%
\newcommand{\xmark}{\ding{55}}%

\def\scititle{
	\textsc{Flowr} -- Flow Matching for Structure-Aware \textit{De Novo}, Interaction- and Fragment-Based Ligand Generation
}
\title{\bfseries \boldmath \scititle}

\author{
    \textbf{Julian Cremer}$^{1,\ast,\dagger}$, 
    \textbf{Ross Irwin}$^{2,3,\ast,\dagger}$, 
    \textbf{Alessandro Tibo}$^{2}$, 
    \textbf{Jon Paul Janet}$^{2}$,\\
    \textbf{Simon Olsson}$^{3}$, 
    \textbf{Djork-Arné Clevert}$^{1}$\\[0.5em]
    \small $^{1}$Machine Learning \& Computational Sciences, Pfizer Worldwide R\&D, Berlin, Germany\\
    \small $^{2}$Molecular AI, Discovery Sciences, R\&D, AstraZeneca, Gothenburg, Sweden\\
    \small \parbox[t]{0.9\linewidth}{\centering $^{3}$Department of Computer Science and Engineering,\\ Chalmers University of Technology and University of Gothenburg,\\ Gothenburg, Sweden}\\[0.5em]
    \small $^{\ast}$Corresponding authors. Email: julian.cremer@pfizer.com, rossir@chalmers.se\\
    \small $^{\dagger}$These authors contributed equally to this work.
}

\begin{document} 

\maketitle

\begin{abstract}
We introduce~\textsc{Flowr}, a structure-based framework for the generation and optimization of three-dimensional ligands.~\textsc{Flowr} integrates continuous and categorical flow matching with equivariant optimal transport, enhanced by an efficient protein pocket conditioning. Alongside~\textsc{Flowr}, we present~\textsc{Spindr}, a curated dataset comprising ligand-pocket co-crystal complexes specifically designed to address existing data quality issues. Empirical evaluations demonstrate that~\textsc{Flowr} surpasses current state-of-the-art diffusion- and flow-based methods in terms of PoseBusters-validity, pose accuracy, and interaction recovery, while offering an inference speedup, achieving up to 70-fold faster performance. In addition, we introduce~\textsc{Flowr.multi}, a highly accurate multi-purpose model allowing for the targeted sampling of ligands that adhere to predefined interaction profiles and chemical substructures for fragment-based design without the need of re-training or any re-sampling strategies. Collectively, our results indicate that~\textsc{Flowr} and~\textsc{Flowr.multi} represent an advancement in AI-driven structure-based drug design, substantially enhancing the reliability and applicability of \textit{de novo}, interaction- and fragment-based ligand generation in real-world drug discovery settings.
\end{abstract}

\section{Introduction}
Structure-based drug discovery (SBDD) is an integrated computational and experimental approach that leverages the three-dimensional structures of biological macromolecules to guide the rational design and optimization of bioactive compounds. By analyzing protein or nucleic acid binding sites, SBDD aims to identify ligands capable of effectively modulating biological functions~\cite{anderson_2003,anderson_2012}. Commonly employed techniques within this paradigm include molecular docking, virtual screening, and structure-guided ligand optimization~\cite{bajorath_2004,wang_2015}. Despite notable successes, traditional SBDD methods face substantial challenges, such as the inherent complexity of molecular interactions, the vastness of chemical space, and difficulties in accurately predicting ligand binding poses and affinities~\cite{ferreira_2015,shoichet_2004}.

Recent advances in deep learning have provided promising avenues to overcome these limitations. Classical computational methods, including molecular docking and virtual screening, typically rely on simplified approximations of molecular interactions and struggle to efficiently explore extensive chemical spaces. In contrast, data-driven deep learning approaches, particularly generative models, have demonstrated potential in capturing complex relationships inherent in distributions of experimentally determined ligand-protein complexes~\cite{folding:alphafold,folding:rosettafold,folding:alphafold3}.

Among generative modeling techniques, diffusion models have emerged as particularly promising tools for \textit{de novo} ligand design. These models employ iterative stochastic processes to progressively refine molecular structures from initial random noise into chemically valid conformations~\cite{model:edm,model:midi,model:eqgatdiff}. By incorporating pocket-specific constraints during generation, diffusion models effectively capture the geometric and chemical subtleties of protein-ligand interactions, addressing the challenge of accurately predicting binding poses while generating diverse sets of ligands~\cite{model:luo2021,model:pocket2mol,model:targetdiff,model:diffsbdd,model:eqgatdiff,model:pilot}.

Nevertheless, existing diffusion-based approaches are not without drawbacks. Their reliance on iterative stochastic sampling can result in molecules exhibiting strained conformations, uncommon substructures, and reduced drug-likeness~\cite{model:pilot}. Additionally, these methods typically suffer from prolonged sampling times compared to alternative generative frameworks~\cite{model:semlaflow}.

Recently, generative flow matching models have emerged as an alternative paradigm, offering substantial improvements in generation efficiency~\cite{cfm:gaussian-fm}. Notably, flow matching approaches employing mini-batch~\cite{cfm:minibatch-ot} and equivariant optimal transport~\cite{cfm:equi-ot} have been proposed, with the latter demonstrating particular efficacy in molecular generation tasks~\cite{model:semlaflow}.

Building upon these insights, we introduce~\textsc{Flowr}, a flow matching model specifically designed for the \textit{de novo} generation of three-dimensional ligands explicitly conditioned on structural constraints. Our framework enables the efficient generation of ligands informed by the geometry of a protein pocket by using a dedicated pocket encoding scheme in contrast to prior works. In addition, we propose~\textsc{Flowr.multi}, a versatile extension capable of multi-purpose conditional generation. This model can efficiently and accurately generate ligands adhering to predefined interaction profiles between ligand atoms and pocket residues, and can design ligands around specific chemical substructures such as scaffolds and functional groups, facilitating scaffold elaboration, scaffold hopping, and fragment-based ligand design - all without requiring model retraining or computationally expensive stabilization techniques during inference as in prior work~\cite{model:diffsbdd}.

However, the evaluation of SBDD methodologies remains challenging, primarily due to inherent data quality concerns and prevalent data leakage issues in widely utilized benchmark datasets~\cite{benchmarking:runs-n-poses,dataset:plinder}. In particular, the commonly employed~\textsc{CrossDocked2020} dataset~\cite{dataset:crossdocked} exhibits substantial limitations for practical drug discovery applications, stemming from its reliance on rigid-pocket cross-docking protocols. Consequently, ligands are artificially constrained into non-cognate binding pockets, causing models trained on such data to internalize biased, flawed, and unrealistic distributions of ligand-pocket interactions.

To address these critical issues, we introduce~\textsc{Spindr}, a high-quality benchmark dataset specifically developed for SBDD, derived from the recently presented~\textsc{Plinder} dataset~\cite{dataset:plinder}. In constructing~\textsc{Spindr}, we implemented an extensive filtering and structural refinement pipeline designed to correct structural defects prevalent in existing datasets~\cite{dataset:pdbbind-opt}, accurately infer protonation states, atomic-resolution protein-ligand interaction profiles, and minimize potential data leakage between training and test sets by maintaining the~\textsc{Plinder} dataset split. 

In summary, we propose the~\textsc{Flowr} model that improves upon existing approaches in both generative quality and computational efficiency. Additionally, our multi-purpose approach,~\textsc{Flowr.multi}, enables the generation of ligands conditioned on specific interaction profiles or chemical substructures, substantially increasing the proportion of ligands closely aligned with reference complexes and enhancing applicability in downstream tasks such as hit expansion, hit-to-lead and lead optimization. Finally, our~\textsc{Spindr} dataset provides a high-quality resource for training and evaluating 3D generative models, addressing limitations---particularly regarding pose quality and data leakage---in currently available datasets.

\section{The~\textsc{Spindr} Dataset}



Modeling interactions between protein pockets and ligands has recently been gaining attention as a method for evaluating the quality of binding poses and designing better small molecule drug candidates~\cite{bouysset_recovery_2024,benchmarking:posecheck}. At the same time questions have been raised about the quality of existing benchmark datasets -- PDBBind~\cite{dataset:pdbbind} has been found to contain covalently bound ligands, missing atoms in pockets, and steric clashes between the pocket and the ligand~\cite{dataset:pdbbind-opt}. \textsc{CrossDocked2020}~\cite{dataset:crossdocked}, another commonly used dataset for pocket-conditioned ligand generative models, is based on the PDBBind General set and is also likely to share many of these structural defects. Additionally, questions have also been raised as to how well temporal data splits, which are commonly used to create benchmark test sets, are able to assess models’ abilities to generalise to unseen data since there are often close structural similarities between complexes in the training and test sets.

To address the issues of data quality and information leakage, and to provide rich, fine-grained information on the interactions between protein pockets and small molecule ligands, we present \textsc{Spindr} (Small molecule Protein Interaction Dataset, Refined). Using the recently proposed \textsc{Plinder} dataset~\cite{dataset:plinder} as a starting point we apply an extensive filtering and processing pipeline to produce a refined set of high-quality structures. Specifically, to create the \textsc{Spindr} dataset, we took the \textsc{Plinder} dataset release 06/2024 (\textsc{Plinder} version 2) and applied the following processing pipeline:
\begin{enumerate}
\item \textbf{Initial filtering}. We remove all \textsc{Plinder} systems which contain more than one ligand or have more than one protein chain in the pocket. We then remove all systems where the ligand is marked as one or more of the following: ‘oligo’, ‘ion’, ‘cofactor’, ‘artifact’, ‘fragment’, ‘covalent’, or ‘other’.

\item \textbf{Structure refinement}. We use Schrodinger protein preparation wizard (which uses the OPLS 2005 forcefield~\cite{tool:opls-2005-ff}) to refine the structure of the remaining systems. These tools perform the following:
\begin{enumerate}
    \item Add missing atoms to partially filled residues in the protein.
    \item Convert some non-standard residue types to standard ones.
    \item Assign protonation states to heavy atoms and add hydrogen atoms to both the protein and ligand.
    \item Infer bonds and formal charges for both the protein and ligand.
    \item Apply local energy minimisation to the protein-ligand complex.
\end{enumerate}

\item \textbf{Infer protein-ligand interactions}. We use ProLIF~\cite{bouysset_prolif_2021} to infer the interactions between the protein and ligand at an atomic resolution, creating a binary matrix of shape $N_{prot} \text{x} N_{lig} \text{x} |S|$, where $N_{prot}$ is the number of atoms in the protein, $N_{lig}$ is the number of atoms in the ligand, and $S$ is the set of possible interaction types. We apply ProLIF with the default settings and infer all supported interaction types, $|S|$ = 13.

\item \textbf{Quality filtering}. We apply a final filtering step and accumulate the processed systems into train, validation and testing splits. Here, we ensure that all systems contain RDKit-valid ligands. We also filter out any system which contains atoms other than \{H, C, N, O, F, P, S, Cl, Se, Br\} and any system with fewer than 5 residues in the pocket. Additionally, we filter out all systems containing NAG ligands since we found these were highly overrepresented which would likely create an unwanted bias for generative models. We also filter out all systems derived from the PDB complex “1mvm” since it contains many small DNA fragments and was not originally filtered by \textsc{Plinder}.

\added{\item \textbf{Data deduplication}. Since existing datasets often contain substantial structural redundancy, we experiment with two data deduplication strategies presented in Appendix~\ref{sec:spire-add}. In practice, we found little empricial difference between the original and deduplicated datasets, despite their significantly smaller size (\~10,000--15,000 fewer complexes), suggesting there is a significant amount of redundancy in the original data. However, we use the non-deduplicated version of \textsc{Spindr} for the remainder of this paper, though we make all three versions publicly available to enable further investigation of deduplication strategies in ligand generation tasks.}
\end{enumerate}

Our final dataset contains 35,666 protein-ligand complexes, making \textsc{Spindr} the largest dataset of high-quality, refined structures derived directly from crystallographic data. In Table~\ref{tab:dataset-comparison} we compare some of the features of \textsc{Spindr} to other commonly used dataset for SBDD and docking. Notably, in addition to the features in Table~\ref{tab:dataset-comparison}, we maintain the same data splits as \textsc{Plinder}. The \textsc{Plinder} splits were carefully selected to minimise data leakage between train and test sets and to ensure test systems were always of high-quality. This careful curation enables realistic assessment of models' generalisability to unseen data, unlike many existing benchmarks which contain substantial train-test data leakage~\cite{dataset:plinder}.


\begin{table*}[t]
    \small
    \caption{\textbf{Feature comparison of \textsc{Spindr} with commonly used datasets.} Overview of the additional features provided by the \textsc{Spindr} dataset compared to datasets commonly used for training generative models for structure-based drug design (SBDD) and docking tasks. Checkmarks (\cmark) indicate presence, crosses (\xmark) indicate absence of each feature.}
    \label{tab:dataset-comparison}
    \centering
    \begin{tabular}{l|cccc}
        \toprule
        \multirow{2}{4em}{Dataset} & Crystal Structure & Energy-Minimised & Explicit & Protein-Ligand \\
        & Complexes & Conformations & Hydrogens & Interactions \\
        \midrule
        \textsc{CrossDocked}   & \xmark & \xmark & \xmark & \xmark \\
        \textsc{PDBBind}       & \cmark & \xmark & \xmark & \xmark \\
        \textsc{Spindr}        & \cmark & \cmark & \cmark & \cmark \\
        \bottomrule
    \end{tabular}
\end{table*}

\section*{\textsc{Flowr} -- Structure-Aware Ligand Generation}

\begin{figure}[t!]
    \centering
    \includegraphics[width=1.0\columnwidth]{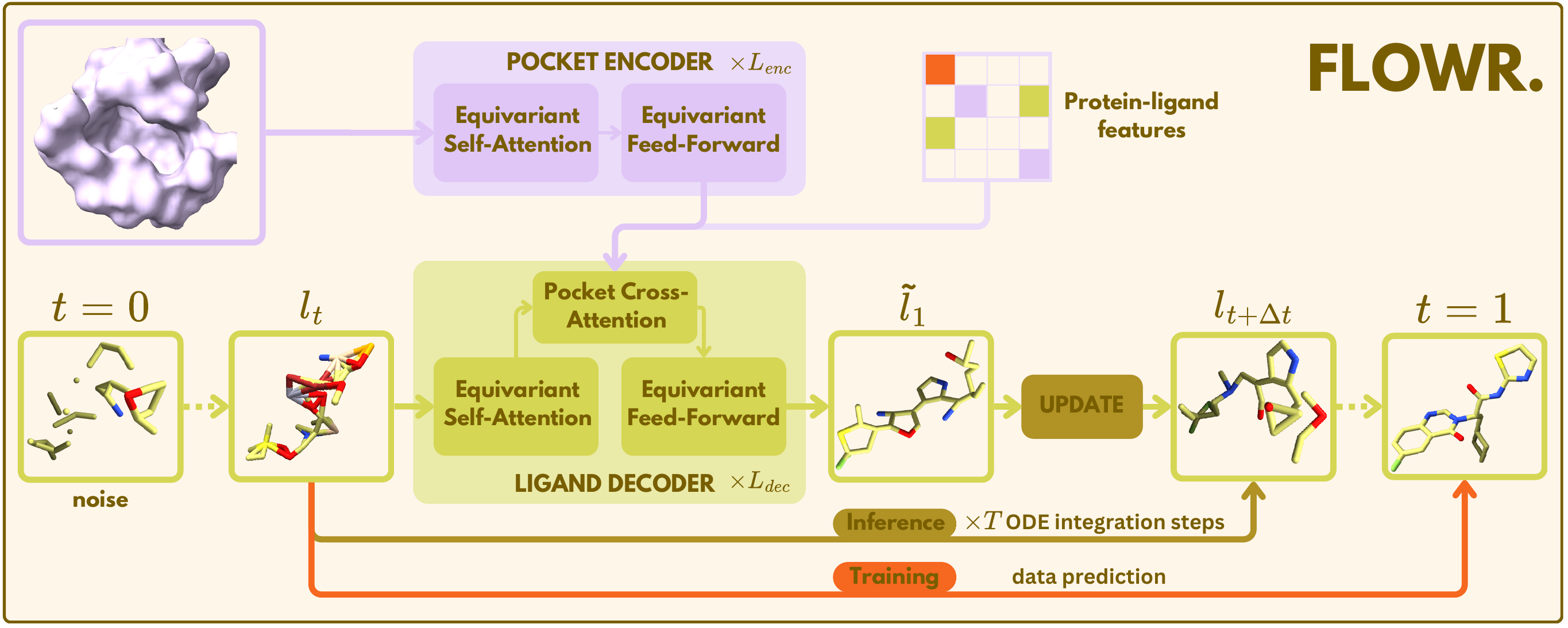}
    \caption{\textbf{Overview of~\textsc{Flowr}.} Schematic overview of the~\textsc{Flowr} model for 3D ligand generation. A protein pocket is encoded and passed, along with the noisy ligand $l_t$, into the ligand decoder, which is trained to produce a denoised ligand $\tilde{l}_t$. Optionally, a set of desired pocket-ligand features can be incorporated. A mixed continuous and categorical flow matching integration scheme is then used to push $l_t$ towards the data distribution and generate a sample $\tilde{l}_1$. The~\textsc{Flowr} model takes as input pocket coordinates along with atom, bond, and residue types, as well as ligand coordinates (with added noise), atom types, and bond types. Pocket features are processed through $L_{enc}$ sequential blocks consisting of equivariant self-attention and equivariant feed-forward layers, resulting in a pocket encoding. This pocket encoding is subsequently integrated via equivariant cross-attention into $L_{dec}$ blocks of equivariant self-attention that process ligand features. Finally,~\textsc{Flowr} outputs denoised ligand coordinates, atom types, bond types, and charges. During inference, the pocket encoding is computed only once and reused for all ligand generation steps. Atom colors: C (yellow), N (blue), O (red), Cl (green).}
    \label{fig:visual-abstract}
\end{figure}


We present \textsc{Flowr} -- a flow-based generative model for \textit{de novo} ligand generation conditioned on a protein pocket and desired pocket-ligand features. We assume access to a dataset containing tuples of a ligand $l$, a protein pocket $\mathcal{P}$ to which the ligand binds, and optionally a matrix $\mathcal{I} \in \mathbb{N}^{M \times N}$ of atomic protein-ligand features, where $M$ and $N$ refer to the number of atoms in the protein and ligand, respectively. In Fig.~\ref{fig:visual-abstract} we show an overview of how our model generates ligands based on protein pocket and pocket-ligand feature conditioning.


The neural network architecture for \textsc{Flowr} is based on the recently proposed \textsc{Semla} architecture~\cite{model:semlaflow}, an E(3)-equivariant message passing framework with latent attention that achieves state-of-the-art results on unconditional 3D molecular generation. We extend \textsc{Semla} to allow conditional generation by incorporating a separate pocket encoder and adding an equivariant cross-attention module within the ligand decoder, enabling structural conditioning on the protein pocket and desired protein-ligand features. Critically, the pocket encoder does not depend on the flow time $t$ or the noisy ligand $l_t$, meaning only one forward pass through the encoder is required when generating ligands, amortising the encoding cost over many samples. We further improve the base architecture by introducing a gated equivariant feed-forward module and passing bond embeddings into every self-attention layer, yielding improved validity and efficiency. Full architectural details and hyperparameters are provided in Appendix~\ref{section:appendix-model}.

\textsc{Flowr} jointly models continuous (coordinates) and discrete (atom types, bond orders) molecular features using a combination of continuous flow matching~\cite{cfm:gaussian-fm,cfm:minibatch-ot} for coordinates and discrete flow models~\cite{model:multiflow} for categorical properties, with equivariant optimal transport~\cite{cfm:equi-ot} to reduce transport costs. Ligand formal charges are directly predicted. The model learns to recover the clean ligand $l_1$ from a noisy interpolant $l_t$ via $p^{\theta}_{1|t} (l_1 | l_t, t; \mathcal{P}, \mathcal{I})$, minimising mean-squared error for coordinates and cross-entropy for categorical features. Given $\mathcal{P}$ and optionally $\mathcal{I}$, ligands are generated by iteratively refining an initial noisy ligand $l_0 \sim p_{\text{noise}}$ through Euler integration of the learned vector field. Full training and sampling details are provided in Appendix~\ref{section:appendix-model}.

\subsection{\textsc{Flowr.multi}: Multi-Purpose Training and Inference}
\label{sec:flowr_multi}

We extend the~\textsc{Flowr} model with~\textsc{Flowr.multi}---a multi-purpose training and inference framework that simultaneously supports both \textit{de novo} generation and any form of fragment-based sampling, like scaffold hopping, scaffold elaboration, fragment linking and fragment-based generation, which is highly relevant from hit expansion to lead optimization campaigns.
As before, we consider a protein pocket $\mathcal{P}$ and a protein-ligand feature matrix $\mathcal{I}$ while assuming a (set of) pre-defined fragmentation(s) applied onto a ligand $l$. Let the ligand consist of $N$ atoms with $\mathbf{l_1} \in \mathbb{R}^{N\times3}$ denoting its coordinates, and, for simplicity, assume that it is split into two fragments containing $n_1$ and $n_2$ atoms, respectively, with $n_1+n_2=N$. Thus, we have $\mathbf{l_{t_1t_2}} = \begin{pmatrix} \mathbf{l^1_{t_1}} \\[1mm] \mathbf{l^2_{t_2}} \end{pmatrix} \in \mathbb{R}^{(n_1+n_2)\times 3}$ with $t_1$ and $t_2$ sampled independently from a uniform distribution. Setting $\mathbf{t_{12}} = \begin{pmatrix} t_1 \\[1mm] t_2 \end{pmatrix}$, the linear interpolation reads $\mathbf{l_{t_1t_2}} = \mathbf{t_{12}} \odot \mathbf{l_1} + (\mathbf{1} - \mathbf{t_{12}}) \odot \mathbf{l_{00}} = \begin{pmatrix} t_1\cdot \mathbf{l^1_{t_1}} + (1-t_1)\cdot \mathbf{l^1_{0}}\\[1mm] t_2 \cdot \mathbf{l^2_{t_2}} + (1-t_2) \cdot \mathbf{l^2_{0}} \end{pmatrix}$, where $\odot$ denotes elementwise multiplication, $\mathbf{1}$ is the all-ones vector, and $\mathbf{l_{00}} \sim p_{\text{noise}}$ is the initial noise sample.

The goal is to learn a joint probability distribution $p^{\theta}_{1|t_{1}t_{2}} (\mathbf{l_1} | \mathbf{l_{t_{1}t_{2}}}, \mathbf{t_{12}}; \mathcal{P}, \mathcal{I})$, from which at inference we sample $\mathbf{\tilde{l}_1} = \begin{pmatrix} \mathbf{\tilde{l}^1_1} \\[1mm] \mathbf{\tilde{l}^2_1} \end{pmatrix} \in \mathbb{R}^{N\times 3}$ to retrieve the joint vector field $f(\mathbf{l_{t_{1}t_{2}}}, \mathbf{t_{12}}; \mathcal{P}, \mathcal{I}) = \mathbf{\tilde{l}_1} - \mathbf{l_{00}}$.

Denoting the per-fragment step sizes by $\Delta t_1 = t_1 + s_1$ and $\Delta t_2 = t_2 + s_2$, where $s_i$ is derived from the number of inference steps, and defining $\mathbf{\Delta t_{12}} = \begin{pmatrix} \Delta t_1 \\[1mm] \Delta t_2 \end{pmatrix}$, the Euler update step reads
\[
\mathbf{l_{t_1+\Delta t_1\ t_2+\Delta t_2}} = \mathbf{l_{t_1t_2}} + \mathbf{\Delta t_{12}}\odot f(\mathbf{l_{t_1t_2}}, \mathbf{t_{12}}; \mathcal{P}, \mathcal{I}) = \begin{pmatrix}
\mathbf{l^1_{t_1}} + \Delta t_1\cdot (\mathbf{\tilde{l}^1_1} - \mathbf{l^1_0})\\[1mm]
\mathbf{l^2_{t_2}} + \Delta t_2\cdot (\mathbf{\tilde{l}^2_1} - \mathbf{l^2_0})
\end{pmatrix}.
\]

Notably, when setting, e.g., $t_1=1$ and $\mathbf{l^1_0} = \mathbf{l^1_{t_1}} = \mathbf{l^1_1}$, we have $\Delta t_1 = 1$ as $s_1$ becomes 0 and the update becomes
\[
\mathbf{l_{1\,t_2+\Delta t_2}} =  \begin{pmatrix}
\mathbf{\tilde{l}^1_1}\\[1mm]
\mathbf{l^2_{t_2}} + \Delta t_2\cdot (\mathbf{\tilde{l}^2_1} - \mathbf{l^2_0})
\end{pmatrix}.
\]

In this scenario, the atoms corresponding to $t_1=1$ remain fixed to be the predictions of the model at each inference step. Assuming the model has successfully learned the identity mapping $\mathbf{\tilde{l}^1_1} = \mathbf{l^1_{1}}$ for the conditional distribution $p^{\theta}_{1|1t{2}} (\mathbf{l_1} | \mathbf{l_{1t_{2}}}, \begin{pmatrix} 1 \\[1mm] t_2 \end{pmatrix} ; \mathcal{P}, \mathcal{I})$, this approach effectively resembles the concept of so-called inpainting. Originally proposed in computer vision~\cite{lugmayr_2022_inpaint}, inpainting has already been adopted for molecular generation tasks~\cite{model:diffsbdd}. However, unlike~\cite{model:diffsbdd}, which requires costly re-sampling steps, and~\cite{model:molsnapper}, which suffers from reduced structural fidelity,~\textsc{Flowr.multi} avoids both limitations while maintaining or even enhancing quality across all generation modes. Specifically, since~\textsc{Flowr.multi} is explicitly trained on a diverse set of inpainting tasks, we anticipate substantial improvements in validity rates, structural accuracy, and inference efficiency, considerably broadening its downstream applicability.

\section{Experiments and Results}

We initially benchmark~\textsc{Flowr} against recent diffusion- and flow-based generative models using the widely adopted~\textsc{CrossDocked2020} dataset~\cite{dataset:crossdocked}. Despite its prevalence, we note that the~\textsc{CrossDocked2020} dataset has several limitations. It is constructed via cross-docking procedures without adequately accounting for pocket flexibility, potentially misrepresenting the natural dynamics and interactions between ligands and their corresponding protein pockets. Moreover, from a practical application standpoint, the ligand chemical space coverage within~\textsc{CrossDocked2020} is limited and notably biased towards non-drug-like or placeholder compounds.

Considering these drawbacks, we shift our primary evaluation to the proposed~\textsc{Spindr} dataset for the remainder of this study. We compare~\textsc{Flowr} directly with~\textsc{Pilot}, a recently proposed diffusion-based model~\cite{model:pilot}.~\textsc{Pilot} has demonstrated marked advancements in distribution learning and ligand quality metrics, outperforming earlier state-of-the-art methods, such as~\textsc{TargetDiff}~\cite{model:targetdiff} and~\textsc{DiffSBDD}~\cite{model:diffsbdd}. Our own evaluations on the~\textsc{CrossDocked2020} dataset confirm~\textsc{Pilot}'s robust performance, establishing it as the strongest competitor and thus the most relevant baseline against~\textsc{Flowr}.
Given~\textsc{Flowr}'s improved computational efficiency and scalability, we further examine the effect of explicitly generating hydrogen atoms in ligands—a critical aspect often overlooked in previous studies despite the fundamental role of hydrogen bonding in protein-ligand interactions.
Crucially, as the~\textsc{Spindr} dataset comes with pre-computed interaction profiles, we also compare~\textsc{Pilot} and~\textsc{Flowr} in terms of interaction recovery, an important metric that helps to better assess ligand quality and distribution learning capabilities~\cite{bouysset_recovery_2024}. We discuss the interactions within the \textsc{Spindr} dataset more in Appendix~\ref{section:appendix-interactions}.

Lastly, we evaluate~\textsc{Flowr.multi}—specifically trained to generate ligands conditioned not only on protein pockets but also on predefined interaction profiles and chemical substructures—on two randomly selected targets from the~\textsc{Spindr} test set: 5YEA and 4MPE. A comprehensive evaluation of~\textsc{Flowr.multi} is provided in Appendix~\ref{app:flowr.multi_eval}.

\subsection{Results}

\begin{table}[t!]
\caption{\textbf{Evaluation and comparison of~\textsc{Flowr} on~\textsc{CrossDocked2020}.} Benchmark comparison of the proposed~\textsc{Flowr} model against~\textsc{Pocket2Mol},~\textsc{TargetDiff},~\textsc{DiffSBDD},~\textsc{Pilot} and~\textsc{DrugFlow} on the~\textsc{CrossDocked2020} test dataset. We follow the conventions in this field and sample 100 ligands per test target, of which there are 100. We evaluate the most expressive metrics, namely PoseBusters-validity, GenBench3D strain energy, AutoDock-Vina scores and the Wasserstein distance of the generated ligands' bond angles (BondA.W1) and bond lengths (BondL.W1) distributions relative to the test set. For all values, we report the mean across ligands and targets and the average standard deviation across targets as subscripts. For all models, we ran all evaluations on the subset of RDKit-valid ligands. Arrows ($\uparrow$/$\downarrow$) indicate that higher or lower values are preferred, respectively.}
\label{tab:results_crossdocked}
\centering
\begin{adjustbox}{width=1.0\textwidth,center}
\begin{sc}
\begin{tabular}{l|c|ccc|cc|cc}
\toprule
Model & PB-valid$\uparrow$ & Strain$\downarrow$ & Vina score$\downarrow$ & Vina score$^{\text{min}}$$\downarrow$ & BondA.W1$\downarrow$ & BondL.W1 [$10^{-2}$]$\downarrow$ & Size & Time (s)$\downarrow$ \\
\toprule
\textsc{\textsc{Pocket2Mol}} & 0.76 {\tiny$\pm$ 0.39} & 147.22 {\tiny$\pm$ 61.41} & -4.72 {\tiny$\pm$ 1.47} & -5.80 {\tiny$\pm$ 1.26} & 2.04 & 0.66 & 17.04 {\tiny$\pm$ 4.11} & 2320 {\tiny$\pm$ 45}\\
\textsc{\textsc{DiffSBDD}} & 0.38 {\tiny$\pm$ 0.46} & 519.03 {\tiny$\pm$ 251.32} & -2.97 {\tiny$\pm$ 5.21} & -4.71 {\tiny$\pm$ 3.30} & 7.00 & 0.51 & 24.85 {\tiny$\pm$ 8.94} & 160.31 {\tiny$\pm$ 73.30}\\
\textsc{\textsc{TargetDiff}} & 0.57 {\tiny$\pm$ 0.46} & 294.89 {\tiny$\pm$ 136.32} & -5.20 {\tiny$\pm$ 1.79} & -5.82 {\tiny$\pm$ 1.60} & 7.76 & 0.42& 22.79 {\tiny$\pm$ 9.46} & 3228 {\tiny$\pm$ 121} \\
\textsc{\textsc{DrugFlow}} & 0.75 {\tiny$\pm$ 0.39} & 120.21 {\tiny$\pm$ 73.28} & -5.66 {\tiny$\pm$ 1.78} & -6.10 {\tiny$\pm$ 1.62} & 2.11 & 0.38 & 21.14 {\tiny$\pm$ 6.81} & - \\
\textsc{Pilot} & 0.83 {\tiny$\pm$ 0.33} & 110.48 {\tiny$\pm$ 87.47} & -5.73 {\tiny$\pm$ 1.72} & -6.21 {\tiny$\pm$ 1.65} & 1.75 & 0.33 & 22.58 {\tiny$\pm$ 9.77} & 295.42 {\tiny$\pm$ 117.35}\\
\midrule
\textsc{Flowr} & 0.92 {\tiny$\pm$ 0.22} & 87.83 {\tiny$\pm$ 74.30} & -6.29 {\tiny$\pm$ 1.56} & -6.48 {\tiny$\pm$ 1.45} & 0.96 & 0.27 & 22.28 {\tiny$\pm$ 9.78} & 12.05 {\tiny$\pm$ 8.01} \\
\toprule
Test set & 0.95 {\tiny$\pm$ 0.21} & 75.62 {\tiny$\pm$ 57.29} & -6.44 {\tiny$\pm$ 2.74} & -6.46 {\tiny$\pm$ 2.61} & - & - & 22.75 {\tiny$\pm$ 9.90} & -\\
\bottomrule
\end{tabular}
\end{sc}
\end{adjustbox}
\end{table}

\begin{figure}[t!]
    \centering
    \includegraphics[width=1.0\columnwidth]{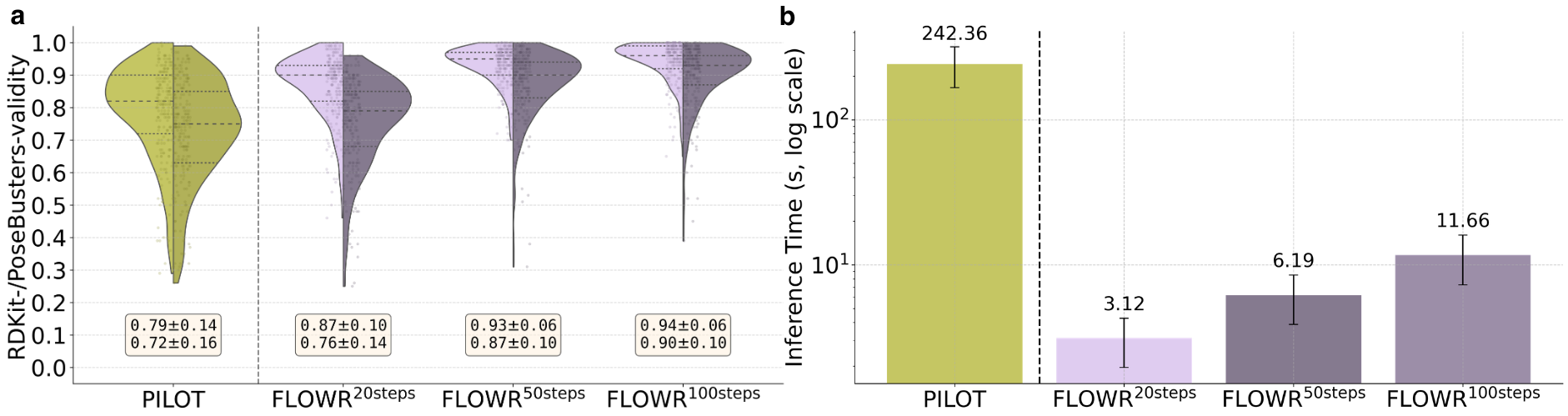}
    \caption{\textbf{Comparison of~\textsc{Pilot} and~\textsc{Flowr} on RDKit-/PoseBusters-validity and inference speed on the~\textsc{Spindr} test set.}
    \textbf{a} Violin plots depicting the distribution of per-target RDKit validity (left half) and PoseBusters (PB) validity (right half) rates across all 225 targets in the~\textsc{Spindr} test set for each model. Each target's validity rate is computed as the fraction of valid molecules out of 100 generated ligands. Dashed horizontal lines indicate the lower quartile, median, and upper quartile of the distribution. Individual per-target rates are overlaid as scatter points. Text annotations below each violin report the mean $\pm$ standard deviation of per-target validity rates (RDkit, PB). \textsc{Flowr} is evaluated at 20, 50, and 100 integration steps. Note, both RDKit- and PoseBusters-validity are evaluated on the full set of generated ligands per target. \textbf{b} Bar plot comparing mean inference times (in seconds, log scale) for generating 100 ligands per target across models. Bar heights represent the mean wall-clock time, with error bars in black indicating the standard deviation. Exact mean values are annotated above each bar. \textsc{Flowr} timings are evaluated at 20, 50, and 100 integration steps. All models are evaluated using a single~\textsc{Nvidia} H100 GPU.}
    \label{fig:timings}
\end{figure}


\begin{figure}[t!]
    \centering
    \includegraphics[width=1.0\columnwidth]{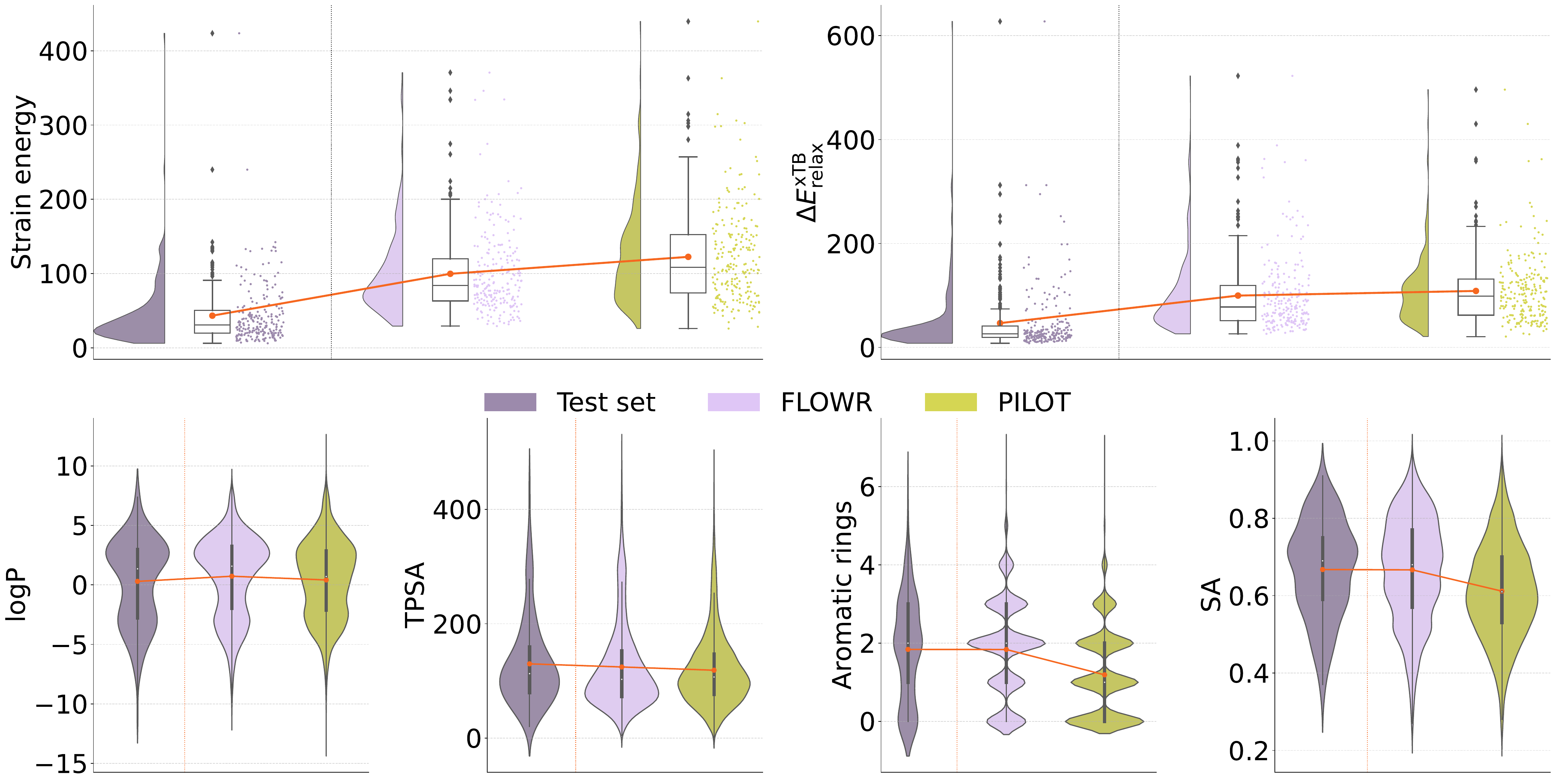}
    \caption{\textbf{Comparison of~\textsc{Pilot} and~\textsc{Flowr} on molecular properties on~\textsc{Spindr}.}
    \textbf{a} Raincloud distribution and box-and-whisker plot showing the distribution of per-target mean strain energy across \textsc{Spindr} test set targets and 100 sampled ligands per target; red connected dots indicate per-model means. \textbf{b} Raincloud distribution and box-and-whisker plot showing \textsc{xTB} relaxation energies ($E_{\text{relax}}^{\text{xTB}}$, kcal/mol) using~\textsc{GFN2-xTB} with implicit solvation and the Analytical Linearized Poisson--Boltzmann (ALPB) solvation model; red connected dots indicate per-model means. Per-target values are computed by averaging over all ligands generated for each target. \textbf{c} Violin plots of per-ligand molecular property distributions for octanol--water partition coefficient (logP), topological polar surface area (TPSA), number of aromatic rings, and synthetic accessibility (SA) score; red connected dots indicate per-model means. Across all panels, test set distributions are shown in purple, \textsc{Flowr} in pink, and \textsc{Pilot} in green. In all box-and-whisker plots, the centre line represents the median, box bounds represent the 25th and 75th percentiles (interquartile range, IQR), and whiskers extend to 1.5$\times$ IQR from the box bounds.}
    \label{fig:mol_properties}
\end{figure}

\begin{figure}[t!]
    \centering
    \includegraphics[width=1.0\columnwidth]{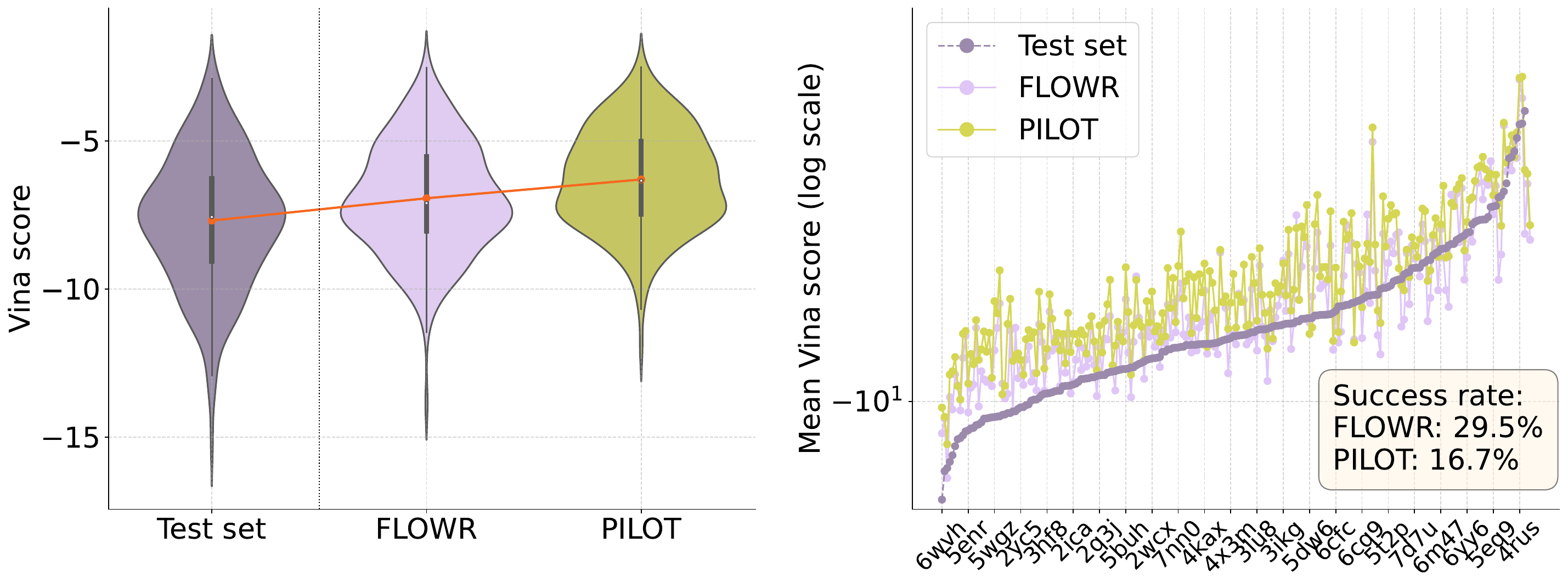}
    \caption{\textbf{Comparison of~\textsc{Pilot} and~\textsc{Flowr} on AutoDock-Vina scores on~\textsc{Spindr}.}
    \textbf{a} Violin plots of per-target mean Vina scores across the \textsc{Spindr} test set targets sampling 100 ligands per target with embedded box plots; red connected dots indicate per-model means. In box plots, the centre line represents the median, box bounds represent the 25th and 75th percentiles (interquartile range, IQR), and whiskers extend to 1.5$\times$ IQR from the box bounds. \textbf{b} Per-target mean Vina scores (symmetric log scale) sorted in ascending order by test set score (purple), with targets labeled every 10th position along the x-axis. \textsc{Flowr} sample scores are denoted by pink, \textsc{Pilot} scores by green scatter points. The inset text box reports the mean per-target success rate, defined as the fraction of generated ligands achieving a Vina score equal to or lower than the corresponding test set mean.}
    \label{fig:vina_scores}
\end{figure}

\begin{table}[t!]
\caption{\textbf{Evaluation and comparison of~\textsc{Pilot} and~\textsc{Flowr} on~\textsc{Spindr}.} Benchmark comparison of the proposed~\textsc{Flowr} model against the~\textsc{Pilot} model using the~\textsc{Spindr} test dataset, which consists of 225 targets. For~\textsc{Flowr}, results are reported for inference steps of 20, 50, and 100. For both models, 100 ligands were sampled per target. The evaluation includes strain energy calculated with GenBench3D and AutoDock-Vina scores (kcal/mol). For all values, we report the mean across ligands and targets and the average standard deviation across targets as subscripts. Additionally, we report the Wasserstein distance of the generated ligands' bond angles (BondA.W1), bond lengths (BondL.W1) and dihedral angles (DihedralW1) distributions relative to those in the~\textsc{Spindr} test set. Ligand sizes for all models are sampled uniformly with a $-$10\%/$+$10\% margin around the respective reference ligand size. Arrows ($\uparrow$/$\downarrow$) indicate that higher or lower values are preferred, respectively.}
\label{tab:results_noHs}
\centering
\begin{adjustbox}{width=1.0\textwidth,center}
\begin{sc}
\begin{tabular}{l|c|c|c|c|c|c}
\toprule
Model & Strain energy $\downarrow$ & Vina score $\downarrow$ & Vina score$^{\text{min}}$ $\downarrow$ & BondA.W1 $\downarrow$ & BondL.W1 [$10^{-2}$] $\downarrow$ & DihedralW1 $\downarrow$ \\
\toprule
\textsc{Pilot} & 120.10 {\tiny$\pm$ 71.61} & -6.30 {\tiny$\pm$ 0.96} & -6.68 {\tiny$\pm$ 1.07} & 1.82 & 0.42 & 5.52 \\
\midrule
\textsc{Flowr}$^{\text{20 steps}}$ & 134.70 {\tiny$\pm$ 77.58} & -6.61 {\tiny$\pm$ 0.98} & -6.92 {\tiny$\pm$ 0.96} & 1.55 & 0.74 & 4.67 \\
\textsc{Flowr}$^{\text{50 steps}}$ & 98.47 {\tiny$\pm$ 56.64} & -6.83 {\tiny$\pm$ 0.93} & -7.13 {\tiny$\pm$ 0.93} & 1.18 & 0.51 & 4.04 \\
\textsc{Flowr}$^{\text{100 steps}}$ & 90.05 {\tiny$\pm$ 52.18} & -6.93 {\tiny$\pm$ 0.92} & -7.22 {\tiny$\pm$ 0.92} & 1.08 & 0.35 & 3.88 \\
\toprule
Test set & 43.27 {\tiny$\pm$ 41.85} & -7.69 {\tiny$\pm$ 2.00} & -7.88 {\tiny$\pm$ 2.00} & - & - & -\\
\bottomrule
\end{tabular}
\end{sc}
\end{adjustbox}
\end{table}

\begin{table}[t!]
\caption{\textbf{Evaluation and comparison of~\textsc{Pilot} and~\textsc{Flowr} on~\textsc{Spindr} with explicit hydrogens.} Benchmark comparison of the proposed~\textsc{Flowr} model against the~\textsc{Pilot} model on the~\textsc{Spindr} test dataset with explicit hydrogens in training and inference. For both models, 100 ligands were sampled per target and evaluated in terms of strain energy, AutoDock-Vina scores (kcal/mol) and Wasserstein distances of generated ligands' bond angle and bond length distributions relative to those in the test set.}
\label{tab:results_withHs}
\centering
\begin{adjustbox}{width=1.0\textwidth,center}
\begin{sc}
\begin{tabular}{l|c|c|c|c|c}
\toprule
Model & Strain energy $\downarrow$ & Vina score$\downarrow$ & Vina score$^{\text{min}}$ $\downarrow$ & BondAnglesW1 $\downarrow$ & BondLengthsW1 [$10^{-2}$] $\downarrow$ \\
\toprule
\textsc{Pilot} & 53.07 {\tiny$\pm$ 22.84} & -5.00 {\tiny$\pm$ 0.65} & -5.50 {\tiny$\pm$ 0.66} & 2.81 & 0.2 \\
\midrule
\textsc{Flowr}$^{\text{100 steps}}$ & 54.11 {\tiny$\pm$ 33.36} & -6.48 {\tiny$\pm$ 0.87} & -6.86 {\tiny$\pm$ 0.87} & 0.82 & 0.1 \\
\toprule
Test set & 43.27 {\tiny$\pm$ 41.85} & -7.69 {\tiny$\pm$ 2.00} & -7.88 {\tiny$\pm$ 2.00} & - & - \\
\bottomrule
\end{tabular}
\end{sc}
\end{adjustbox}
\end{table}

We compare our model against recently published generative methods on the commonly used~\textsc{CrossDocked2020} dataset as an initial benchmark. As can be seen in Tab.~\ref{tab:results_crossdocked}, the proposed~\textsc{Flowr} model substantially outperforms all baseline methods (\textsc{Pocket2Mol}~\cite{model:pocket2mol},~\textsc{DiffSBDD}~\cite{model:diffsbdd},~\textsc{TargetDiff}~\cite{model:targetdiff},~\textsc{DrugFlow}~\cite{drugflow}, and~\textsc{Pilot}~\cite{model:pilot}) across the evaluated metrics on the~\textsc{CrossDocked2020} test dataset. Specifically,~\textsc{Flowr} achieves the highest PoseBusters-validity (0.92 ± 0.22), lowest strain energy (87.83 ± 74.30), best AutoDock-Vina scores (mean: -6.29 ± 1.56, minimized: -6.48 ± 1.45), and lowest Wasserstein distances for bond angles (0.96) and bond lengths (0.27). Additionally,~\textsc{Flowr} demonstrates substantially faster inference time (12.05 ± 8.01 seconds) compared to other methods. These results indicate that~\textsc{Flowr} generates ligand conformations closest to the test set distribution and with superior computational efficiency. We note that the second best model,~\textsc{Pilot}, also shows substantially better results compared to all other methods, especially in terms of PoseBusters-validity (0.83 ± 0.33). Thus, we selected~\textsc{Pilot} as our main competitor for the remaining of this study.

In Fig.~\ref{fig:timings}, we compare~\textsc{Pilot} and~\textsc{Flowr} trained on the~\textsc{Spindr} training dataset in terms of RDKit-validity, PoseBusters-validity (PB-validity), and inference speed on the~\textsc{Spindr} test set. Our results indicate that~\textsc{Flowr} generates ligands with substantially higher validity on average. While RDKit-validity is a 2D ligand-centric measure, the PoseBusters suite~\cite{benchmarking:posebusters} evaluates ligand conformations using well-established 3D ligand-pocket-based metrics, providing a more comprehensive assessment of pose accuracy.~\textsc{Flowr} achieves a substantial improvement over~\textsc{Pilot} in both metrics, with an average RDKit-validity of 0.94 ± 0.24 vs. 0.79 ± 0.39 and an average PB-validity of 0.88 ± 0.21 vs. 0.71 ± 0.18, respectively.
Notably,~\textsc{Flowr} substantially improves inference speed, outperforming~\textsc{Pilot} by a factor of 20 when using 100 inference steps, as shown in Fig.~\ref{fig:timings} (right). This efficiency gain is primarily attributed to~\textsc{Flowr}'s model architecture and the protein pocket encoder requiring only a single forward when integrating the vector field. In contrast, prior models~\cite{model:targetdiff, model:diffsbdd, model:eqgatdiff, model:pilot} often recompute protein pocket embeddings at every sampling step. Notably, the number of integration steps can be reduced as low as 20, achieving a 70-fold speed-up over~\textsc{Pilot} while impacting model performance comparably little.

In Tab.~\ref{tab:results_noHs} we compare~\textsc{Pilot} and~\textsc{Flowr} in terms of strain energy (calculated using GenBench3D~\cite{benchmarking:genbench3d}), AutoDock-Vina score (used as an approximate measure of pose quality and binding affinity~\cite{docking:autodock-vina}), and their ability to generalize to the test set distribution based on Wasserstein distance measures for bond angles, bond lengths and dihedral angles following~\cite{model:midi, model:eqgatdiff, model:pilot}. A more comprehensive overview of results is given in Tab.~\ref{tab:full_results} in the Appendix. As flow matching allows for setting the number of inference steps, we also report the same results for different number of steps, namely 20, 50 and 100 (default).

In terms of strain energy values,~\textsc{Flowr} substantially outperforms~\textsc{Pilot} (90.05 ± 52.18 vs. 120 ± 71.61). However, we note that, on average, the strain energies of generated ligands do not align well with those of the test set (43.27 ± 41.85), as illustrated in Fig.~\ref{fig:mol_properties} (top left). We hypothesize that this discrepancy arises primarily from limited coverage of chemical and conformational space in the training data, due to the relatively low availability of co-crystal structures. \added{Notably, these elevated strain energies reflect subtle deviations in bond angles and lengths rather than gross structural defects—simple MMFF94s-based relaxation with fixed protein pockets reduces strain energies to 28.05 ± 28.25 kcal/mol, below the test set average. In addition, a mean RMSD of 0.775 ± 0.144~\r{A} confirms that generated molecules readily relax to low-energy conformations with minimal structural perturbation (see Fig.~\ref{fig:strain_analysis} in App.~\ref{sec:appendix-results}), while mean PB-validity increases to 0.95 ± 0.08 and mean Vina score decreases to -6.97 ± 0.89.} Additionally, Fig.~\ref{fig:mol_properties} (top right) shows the relaxation energy distribution of generated ligands calculated using the~\textsc{GFN2-xTB} method~\cite{xtb_gfn2} with implicit solvation using the~\textsc{ALPB} model~\cite{alpb} (46.37 ± 64.05 kcal/mol on the test set; 100.37 ± 59.85 for~\textsc{Flowr} vs. 107.89 ± 66.37 for~\textsc{Pilot}). Another commonly reported metric in this context is the clash count between ligand and pocket atoms. Using PoseCheck~\cite{benchmarking:posecheck}, we observe a clear improvement for~\textsc{Flowr} (5.25 ± 2.21) compared to the~\textsc{Pilot} model (6.28 ± 2.61), with~\textsc{Flowr} more closely resembling the test set distribution (4.24).

\textsc{Flowr} outperforms~\textsc{Pilot} in docking assessments, suggesting a higher pose accuracy (-6.93 ± 0.92 vs. -6.30 ± 0.96). We use Vina's scoring function with no re-docking applied, but also report the minimized Vina score, where local energy minimization is applied to the ligand (-7.22 ± 0.92 vs. -6.68 ± 1.07). In Fig.~\ref{fig:vina_scores} (left) we compare the Vina score distribution across targets and the mean Vina score per target (right, log-scale). As can be seen,~\textsc{Flowr} shows a 12.8\% increase in success rate (number of ligands per target that are either equal or better than the reference with respect to Vina scoring) with an average success of 29.5\%.

Additionally, we measure distribution learning capabilities in terms of bond angle and bond length Wasserstein distances to the test set. Here,~\textsc{Flowr} demonstrates substantially better generalization compared to~\textsc{Pilot} with a mean bond angles distance of 1.08 vs. 1.82, mean bond lengths distance of 0.35 vs. 0.42 and mean dihedral angles distribution distance of 5.51 vs 3.45.
We observe similar results when comparing both models on a set of relevant molecular properties like lipophilicity (logP), topological polar surface area (TPSA), number of aromatic rings and the synthesizability of generated molecules against the test set shown in Fig.~\ref{fig:mol_properties} (bottom). While~\textsc{Pilot} shows similar distributions for both logP and TPSA, it is substantially worse in reproducing the number of aromatic rings and similarly synthesizable compounds compared to the test set. In Fig.~\ref{fig:flowr_pilot_walters} in the Appendix, we provide additional results comparing~\textsc{Flowr} and~\textsc{Pilot} on the~\textsc{Spindr} test set using key drug-likeness filters proposed by~\cite{walters_filter}\footnote{Inspired by \href{http://practicalcheminformatics.blogspot.com/2024/05/generative-molecular-design-isnt-as.html}{this blog post} by Pat Walters.}. These findings demonstrate that ligands generated by~\textsc{Pilot} exhibit up to 40\% lower pass-through rates compared to those generated by~\textsc{Flowr}, with~\textsc{Flowr}'s results aligning substantially more closely with the~\textsc{Spindr} test data.

In Tab.~\ref{tab:results_withHs}, we repeat the same experiments while incorporating explicit hydrogens in the ligands for both training and inference. Under these conditions,~\textsc{Pilot} exhibits a clear decrease in performance, while~\textsc{Flowr} maintains comparable results. However, for both models validity drops substantially, with RDKit-validity decreasing to 0.64 ± 0.48 for~\textsc{Flowr} and 0.52 ± 0.50 for~\textsc{Pilot}, while PB-validity declines to 0.60 ± 0.22 and 0.47 ± 0.14, respectively.
Notably, strain energy metrics improve substantially with explicit hydrogen modeling for both models. While heavy-atom-only approaches require post-hoc hydrogen addition potentially leading to artificially inflated strain energies, analysis of individual PoseBusters metrics reveals that reduced validity stems predominantly from ligand-protein clashes rather than internal molecular geometry issues. Since~\textsc{Spindr} provides limited coverage of both chemical and conformational space, we anticipate these clashes will diminish with increased training data and, critically, by modeling explicit hydrogens also in protein pockets alongside ligands.

Overall, the proposed~\textsc{Flowr} model consistently outperforms~\textsc{Pilot} across all evaluated metrics, demonstrating substantially improved capability in modeling and generalizing ligand-pocket complex distributions. Specifically, we observe an average increase of approximately 15\% in ligand and ligand-pocket validity metrics, along with substantially improved AutoDock-Vina scores, indicating higher-quality generated poses. Interestingly, using~\textsc{Flowr} with only 50 inference steps consistently yields better results and yields comparable or slightly worse results with 20 inference steps compared to the~\textsc{Pilot} model with 500 steps. Thus,~\textsc{Flowr} achieves substantial performance gains while reducing inference time by up to a factor of 70.

Nevertheless, there remains room for improvement, particularly in reducing strain energies of generated ligands. Additionally, accurately modeling ligand-pocket complexes with explicit hydrogens continues to be challenging, especially in scenarios with limited training data. We encourage the scientific community to evaluate generative models incorporating explicit hydrogen atoms as a challenging benchmark in future research. In this context, the proposed~\textsc{Spindr} dataset represents a valuable resource, providing a robust and comprehensive benchmark for evaluating and comparing ligand generation models.

\subsubsection{Interaction recovery}

\begin{figure}[t!]
    \centering
    \includegraphics[width=1.0\columnwidth]{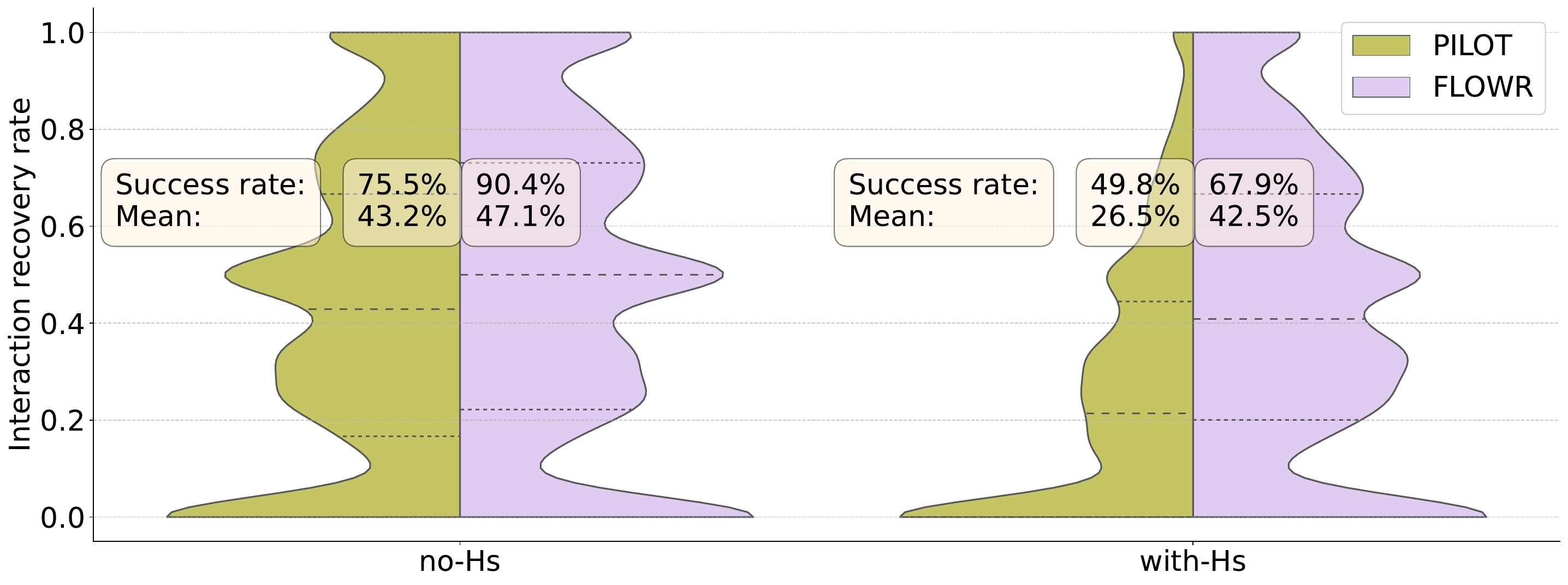}
    \caption{\textbf{Comparison of~\textsc{Pilot} and~\textsc{Flowr} on interaction recovery on~\textsc{Spindr}.}
    Comparison of~\textsc{Pilot} and~\textsc{Flowr} in terms of interaction recovery rate distributions across the \textsc{Spindr} test set targets sampling 100 ligands per target, shown as violin plots with embedded box plots. Both models are either trained without explicit hydrogens (no-Hs) or with explicit hydrogens (with-Hs). The success rate is the percentage of ligands for which interaction fingerprints could be retrieved. Dashed lines indicate the lower quartile (25th percentile), median, and upper quartile (75th percentile). Green color denotes \textsc{Pilot} and pink \textsc{Flowr} distributions. Interaction fingerprints were calculated using~\textsc{ProLIF}.}
    \label{fig:interaction_recovery}
\end{figure}

In SBDD, understanding how a ligand interacts with its target binding site at the atomic level is essential for optimizing potency, selectivity, and pharmacological properties~\cite{salentin_plip_2015, jubb_arpeggio_plip_2017, bouysset_prolif_2021}. Ligand-pocket interactions—including hydrogen bonds, hydrophobic contacts, $\pi$–$\pi$ and $\pi$–cation stacking, salt bridges, and electrostatic or van der Waals interactions—collectively determine binding affinity and specificity. Consequently, these protein-ligand interactions, or more precisely, the ligand's binding pose, are crucial for assessing biological relevance and activity~\cite{bouysset_recovery_2024}. To systematically identify such interactions, protein-ligand interaction fingerprints (PLIFs) are commonly employed~\cite{bouysset_prolif_2021, bouysset_recovery_2024}. PLIFs encode key interaction features, including the interacting protein residue, interaction type, and optionally, the ligand atom involved~\cite{bouysset_prolif_2021, bouysset_recovery_2024}.

In contrast to previous studies, we closely investigate our proposed model's capability to reproduce interactions observed in reference ligands. Figure~\ref{fig:interaction_recovery} illustrates the distribution of interaction recovery rates for~\textsc{Pilot} and~\textsc{Flowr} across the~\textsc{Spindr} test set targets, both with and without explicit hydrogen modeling, using the same sampling settings as before. We also report the success rate, defined as the proportion of RDKit- and PoseBusters-valid ligands for which interactions could be identified. As shown,~\textsc{Flowr} consistently outperforms~\textsc{Pilot} (47.1\% vs. 43.2\%, with success rates of 90.4\% vs. 75.5\%), particularly when explicitly modeling hydrogens (42.5\% vs. 26.5\%, with success rates of 67.9\% vs. 49.8\%).

However, these results suggest that a purely \textit{de novo} generation approach may be less suitable for targeted ligand generation tasks commonly employed in hit expansion and optimization campaigns. To address this, we propose using~\textsc{Flowr.multi} (Sec.~\ref{sec:flowr_multi}), a multi-purpose model capable of interaction-conditional generation. We present detailed results for this approach in the following section.

\subsection{\textsc{Flowr.multi}: Interaction-conditional generation}

\begin{figure}[t!]
    \centering
    \includegraphics[width=1.0\columnwidth]{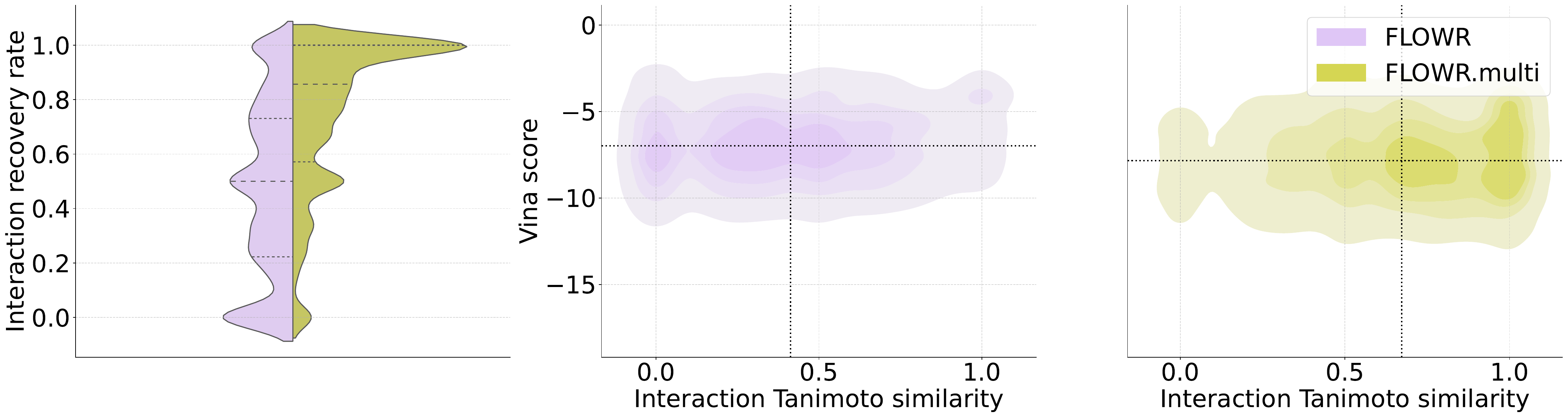}
    \caption{\textbf{Comparison of~\textsc{Flowr} and~\textsc{Flowr.multi} on interaction similarity and Vina scores on~\textsc{Spindr}.}
    \textbf{a} Split violin plot of per-ligand interaction recovery rates across the \textsc{Spindr} test set targets with 100 sampled ligands per target. Dashed lines indicate the lower quartile (25th percentile), median, and upper quartile (75th percentile); the left half corresponds to \textsc{Flowr} (pink) and the right half to \textsc{Flowr.multi} (green). \textbf{b,c} Two-dimensional (2D) kernel density estimates of per-ligand protein--ligand interaction fingerprint (PLIF) Tanimoto similarity versus AutoDock Vina docking score for \textsc{Flowr} (\textbf{b}) and \textsc{Flowr.multi} (\textbf{c}). Dotted lines indicate the respective mean values. Interaction fingerprints were calculated using~\textsc{ProLIF}.}
    \label{fig:flowr_vs_flowr-cond}
\end{figure}

To improve interaction recovery, we propose to use~\textsc{Flowr.multi} with an interaction-based fragmentation for training and inference, which ensures that in inpainting-mode ligand atoms involved in pocket interactions are kept fixed. Let $\mathbf{X}_p = \{\mathbf{x}_{p,j} \in \mathbb{R}^3 : j=1,\dots,n_p\}$ denote the 3D coordinates of the $n_p$ pocket atoms, $\mathbf{X}_l = \{\mathbf{x}_{l,i} \in \mathbb{R}^3 : i=1,\dots,n_l\}$ denote the ground-truth (native) 3D coordinates of the $n_l$ ligand atoms, $I \in \{0,1\}^{n_p \times n_l \times d_I}$ be an interaction tensor, where the entry $I_{j,i,k}$ indicates whether pocket atom $j$ and ligand atom $i$ participate in an interaction of type $k$ (with $d_I$ possible interaction channels). We define a binary mask $M\in\{0,1\}^{n_l}$ by
\begin{align}M_i = \mathbb{I}\Biggl\{ \sum_{j=1}^{n_p}\sum_{k=1}^{d_I} I_{j,i,k} > 0 \Biggr\}, \quad i=1,\dots,n_l, \end{align}
where $\mathbb{I}\{\cdot\}$ denotes the indicator function. This mask partitions the ligand atoms into a set of interacting atoms (superscript $I$) for which we fix the flow time to $t_I=1$ and set the noise to the ground truth $\mathbf{l}^I_0 = \mathbf{l}^I_1$, and a set of remaining atoms that are generated unconstrained, as described in Sec.~\ref{sec:flowr_multi}.

Using~\textsc{Flowr.multi}, we achieve an average interaction recovery rate of 76.1\%; the distribution compared to the~\textsc{Flowr} model is shown in Fig.~\ref{fig:flowr_vs_flowr-cond} (left). Notably, despite the conditional generation process, the model maintains its ability to explore chemical space, achieving an average molecular diversity of 0.83 compared to 0.86 for~\textsc{Flowr}. As illustrated in Fig.~\ref{fig:flowr_vs_flowr-cond} (right),~\textsc{Flowr.multi} also substantially improves predicted binding affinity, as indicated by a lower average Vina score (-7.18 vs. -6.93), while interaction Tanimoto similarity nearly doubles.

Thus, we observe that~\textsc{Flowr.multi} effectively generates ligands adhering to predefined interaction profiles, improving pose accuracy (as measured by Vina scoring) without substantially compromising chemical diversity or exploration compared to the purely~\textit{de novo}~\textsc{Flowr} model. For a more comprehensive overview of the performance of~\textsc{Flowr.multi} on the~\textsc{Spindr} test dataset for interaction-conditional and multi-purpose generation, we refer to Tab.~\ref{tab:flowr.multi-results} in the Appendix.
In the next section, we further investigate the multi-purpose capabilities of~\textsc{Flowr.multi} in more details using two test set targets.

\subsection{\textsc{Flowr.multi}: Multi-purpose generation on 5YEA and 4MPE}

\begin{table}[t!]
    \caption{\textbf{Evaluation of~\textsc{Flowr.multi} on 5YEA and 4MPE.}
    Performance evaluation for interaction-, scaffold-, and functional group-conditional generation with~\textsc{Flowr.multi} on two randomly selected test targets with PDB-ID 5YEA and 4MPE, respectively. We report PoseBusters-validity (PB-validity) across 100 ligands per target, the mean Vina score (kcal/mol) as well as interaction recovery rate (PLIF recovery) and synthesizability score (SA score). }
    \label{tab:case_study}
    \centering
    \begin{adjustbox}{width=1.0\textwidth,center}
    \begin{sc}
    \label{tab:protein_results}
    \begin{tabular}{l|l|c|ccc}
        \toprule
        Protein & Metric & Reference &~\textsc{Flowr.multi}$^{\mathrm{interact.\text{-}cond.}}$ &~\textsc{Flowr.multi}$^{\mathrm{scaffold\text{-}cond.}}$ &~\textsc{Flowr.multi}$^{\mathrm{f.\,group-\,cond.}}$ \\
        \midrule
        \multirow{4}{*}{5YEA} & PB-validity $\uparrow$  & 1.0 & 0.90  & 0.98  & 0.89  \\[0.5ex]
                             & Vina score $\downarrow$    & -9.57 & -8.96  & -8.71  & -8.99  \\[0.5ex]
                             & Vina score (Top-10) $\downarrow$ & - & -10.08 & -10.16 & -8.99 \\[0.5ex]
                             & PLIF recovery rate $\uparrow$ & - & 0.87  & 0.75  & 0.77  \\[0.5ex]
                             & SA score $\uparrow$      & 0.82 & 0.77 & 0.82 & 0.76 \\
        \midrule
        \multirow{4}{*}{4MPE} & PB-validity $\uparrow$  & 1.0 & 0.95 & 1.0 & 0.92 \\[0.5ex]
                             & Vina score $\downarrow$    & -7.23 & -6.80 & -7.27 & -6.41 \\[0.5ex]
                             & Vina score (Top-10) $\downarrow$ & - & -7.54 & -7.83 & -7.15 \\[0.5ex]
                             & PLIF recovery rate $\uparrow$ & - & 0.79 & 0.53 & 0.89 \\[0.5ex]
                             & SA score $\uparrow$      & 0.84 & 0.81 & 0.82 & 0.82 \\
        \bottomrule
    \end{tabular}
    \end{sc}
    \end{adjustbox}
\end{table}

\begin{figure}[t!]
    \centering
    \includegraphics[width=1.0\textwidth]{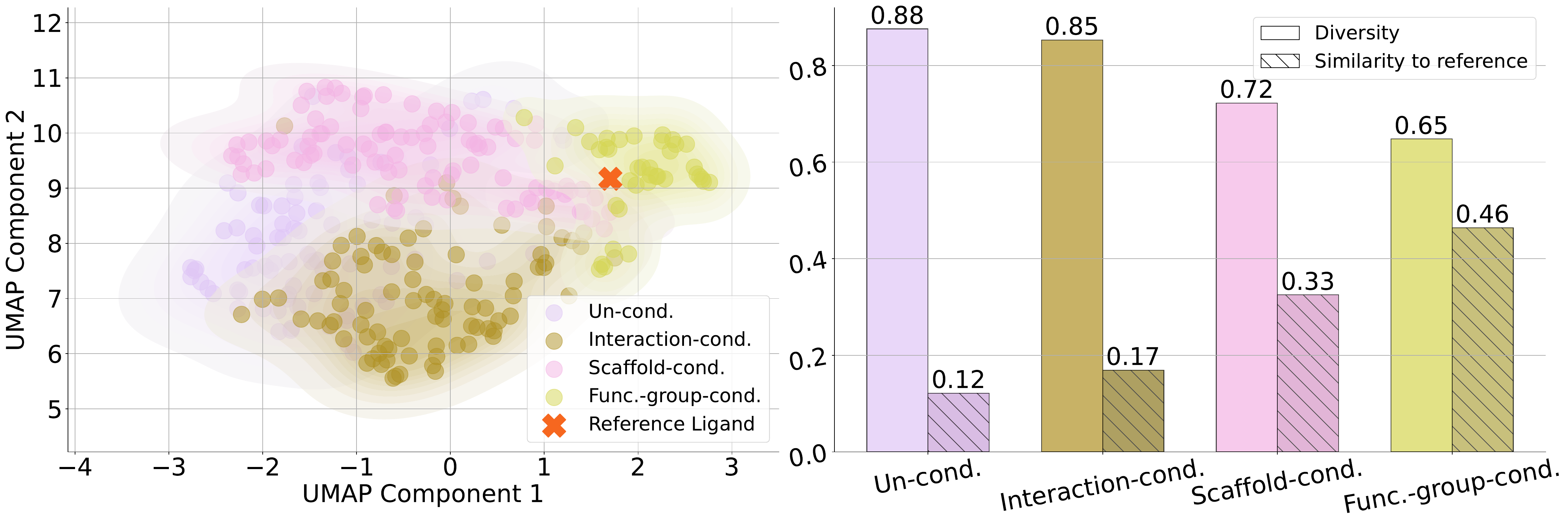}
    \caption{\textbf{Evaluation of chemical space coverage on 5YEA with~\textsc{Flowr.multi}.}
    Using the unconditional, interaction-, scaffold-, and functional group-conditional generation modes of~\textsc{Flowr.multi}, 100 ligands were sampled for a randomly selected target from the~\textsc{Spindr} test set (here: PDB-ID 5YEA). Chemical space coverage is visualized with respect to the reference ligand using Morgan fingerprints and Uniform Manifold Approximation and Projection (UMAP). Average diversity of the sampled ligand sets and average similarity to the reference are reported. Each generation mode is represented by a distinct color as indicated in the legend.}
    \label{fig:flowr_umap_5yea}
\end{figure}

\begin{figure}[t!]
    \centering
    \includegraphics[width=1.0\textwidth]{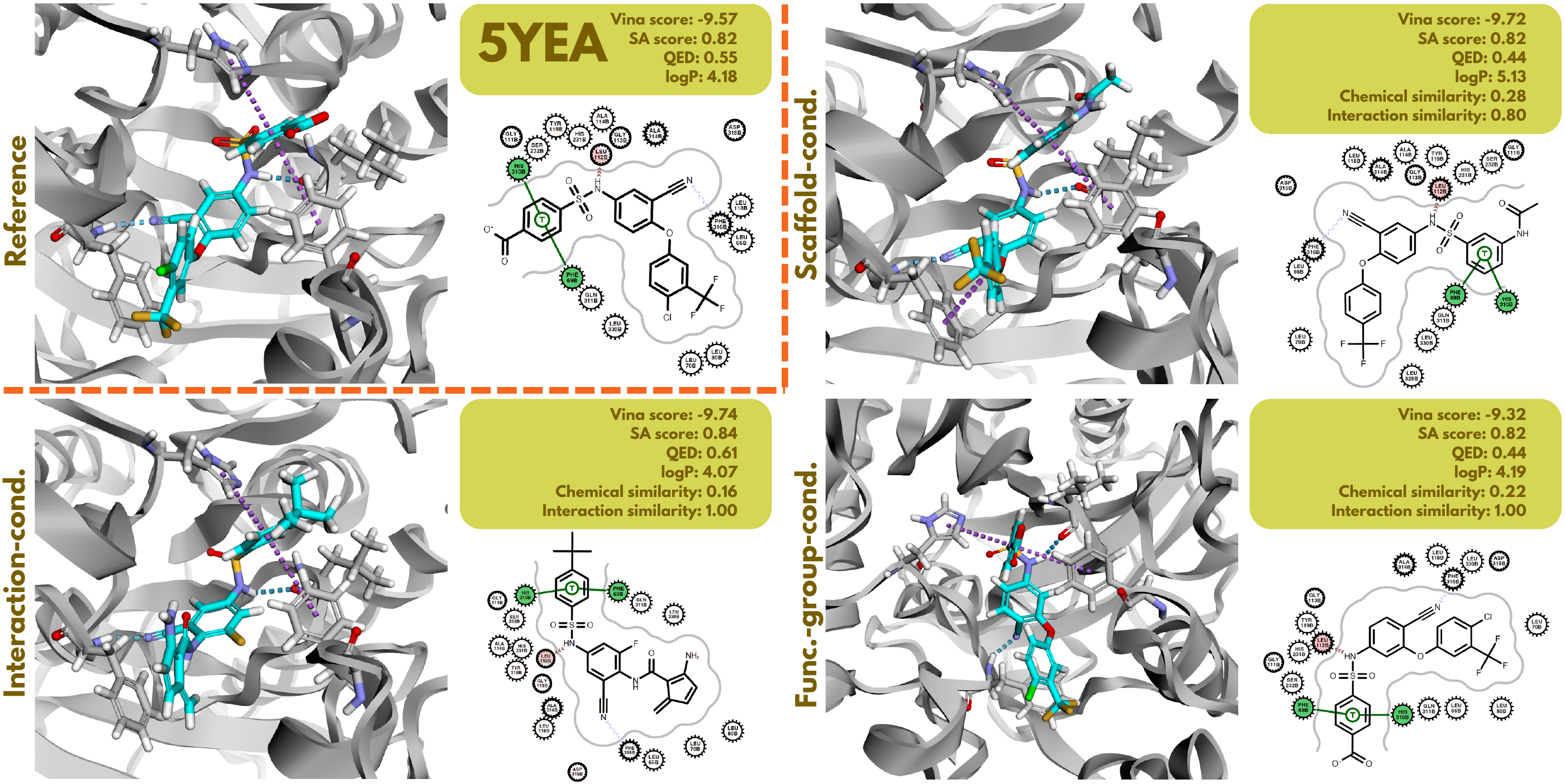}
    \caption{\textbf{Visualization of conditional generation on 5YEA with~\textsc{Flowr.multi}.}
    Using the interaction-, scaffold-, and functional group-conditional generation modes of~\textsc{Flowr.multi}, 100 ligands for a randomly selected target from the~\textsc{Spindr} test set (here: PDB-ID 5YEA) were sampled. Three ligands were selected at random and compared to the reference compound based on Vina score, synthetic accessibility (SA) score, quantitative estimate of drug-likeness (QED), octanol--water partition coefficient (logP), chemical similarity, and interaction similarity. Atom colors: C (cyan/gray), N (blue), O (red), S (yellow), F (ochre), Cl (green), H (white). Interaction fingerprints were calculated using~\textsc{ProLIF}. Interaction diagrams were generated using the \textsc{OpenEye Python Toolkit}.}
    \label{fig:flowr_cond_5yea}
\end{figure}

\begin{figure}[t!]
    \centering
    \includegraphics[width=1.0\textwidth]{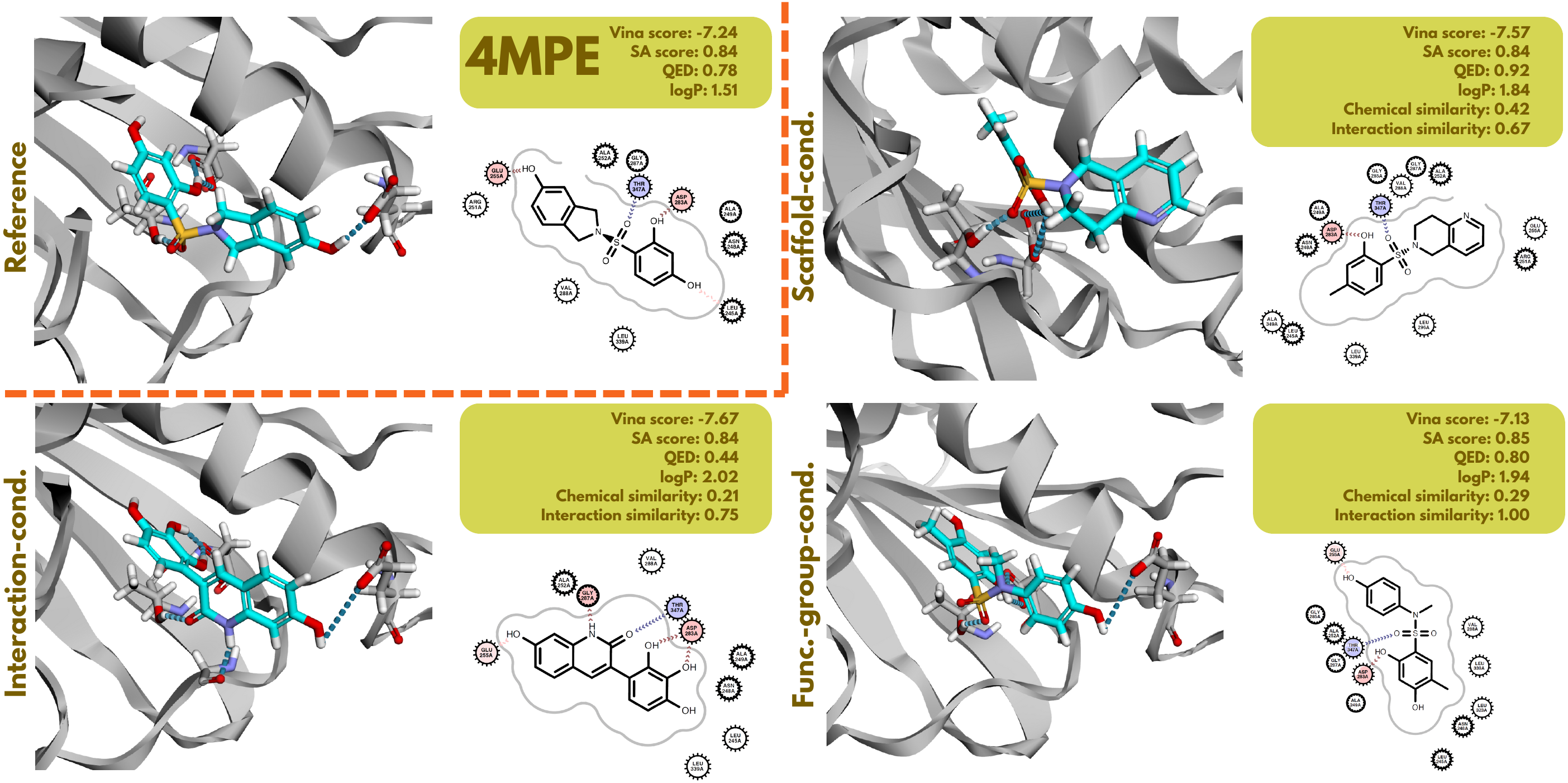}
    \caption{\textbf{Evaluation of conditional generation on 4MPE with~\textsc{Flowr.multi}.}
    Using the interaction-, scaffold-, and functional group-conditional generation modes of~\textsc{Flowr.multi}, 100 ligands for a randomly selected target from the~\textsc{Spindr} test set (here: PDB-ID 4MPE) were sampled. Three ligands were selected at random and compared to the reference compound based on Vina score, synthetic accessibility (SA) score, quantitative estimate of drug-likeness (QED), octanol--water partition coefficient (logP), chemical similarity, and interaction similarity. Atom colors: C (cyan/gray), N (blue), O (red), S (yellow), F (ochre), Cl (green), H (white). Interaction fingerprints were calculated using~\textsc{ProLIF}. Interaction diagrams were generated using the \textsc{OpenEye Python Toolkit}.}
    \label{fig:flowr_three_cond_4mpe}
\end{figure}

To evaluate the multi-purpose generative capabilities of the~\textsc{Flowr.multi} model, we randomly selected two targets (PDB IDs: 5YEA and 4MPE) from the test set and generated ligands under three distinct conditions: interaction-conditional, scaffold-conditional, and functional group-conditional generation. Note that \textsc{Flowr.multi} can also be applied to tasks such as fragment linking and fragment growing; however, for clarity, we leave the evaluation of these additional applications to future work.

The selected crystal structure 5YEA represents lipoprotein-associated phospholipase A2 (Lp-PLA2), a validated therapeutic target implicated in atherosclerosis, Alzheimer's disease, and diabetic macular edema~\cite{5yea}. Potent inhibitors has been found through fragment screening, molecular docking, and structure-guided optimization, achieving substantial potency improvements from micromolar to single-digit nanomolar inhibitors~\cite{5yea}. Given the proven effectiveness of structure-based approaches for this target, applying~\textsc{Flowr.multi} to Lp-PLA2 (5YEA) is particularly interesting.
The other selected protein target, 4MPE, corresponds to pyruvate dehydrogenase kinase (PDK), an enzyme family (isoforms 1–4) that negatively regulates mitochondrial pyruvate dehydrogenase complex activity through phosphorylation~\cite{4mpe}. PDK isoforms are clinically relevant, as their overexpression is associated with obesity, diabetes, heart failure, and cancer, making them attractive therapeutic targets. Previous works explored a structure-guided approach to design selective inhibitors targeting the conserved ATP-binding pocket of PDK isoforms, resulting in the potent inhibitor PS10~\cite{4mpe}, making it an interesting reference point for our~\textsc{Flowr.multi} model.

For each target and condition, we generated 100 ligands and compared the resulting ligand distributions to the respective reference ligand in terms of PoseBusters-validity, Vina docking score, interaction recovery rate, and synthetic accessibility. The results are summarized in Table~\ref{tab:case_study}. We consistently observe high PoseBusters-validity across targets, suggesting that~\textsc{Flowr.multi} effectively learned to generate accurate ligand poses independent of the conditioning mode. Although the mean Vina scores across generated ligands does not match those of the reference ligands, selecting the top-10 ligands based on Vina scores consistently yielded ligands with superior docking scores compared to the references (with slightly worse results for functional group-conditional generation). Interaction recovery rates are generally close to 1, indicating that the generated ligands closely reproduce the interaction profiles of the reference ligands. Finally, the mean SA scores, indicative of synthesizability, are consistently around or above 0.80, comparable to the reference ligands. This suggests that the generated ligands not only satisfy relevant physicochemical criteria but are also likely to be synthetically accessible.
In Fig.~\ref{fig:flowr_umap_5yea} we show the chemical space coverage of generated ligands per generation mode, evaluate the sample diversity and the diversity towards the reference compound. Notably, we find a strong dependence of ligand diversity and reference similarity on the condition-mode. While in the~\textit{de novo} setting we get the most diverse set of ligands, as to be expected, the interaction-conditional also shows a strong chemical space coverage although interaction recovery is substantially enhanced reaching almost 90\%. However, especially the functional-group-conditional setting allows for a close resemblance of the reference's chemical space. This is interesting as this shows that practitioners can use different conditional setups of~\textsc{Flowr.multi} for controlled chemical space exploration.
In Fig.~\ref{fig:flowr_cond_5yea} we visualize a randomly selected ligand for 5YEA per conditioning mode and compare to the reference ligand. Fig.~\ref{fig:flowr_three_cond_4mpe} shows the corresponding visualization for 4MPE. More examples and visualizations for both targets are provided in Appendix~\ref{sec:appendix-results}.

\FloatBarrier
\subsection{Conclusion}

In this work, we introduced~\textsc{Flowr}, a generative framework for structure-based \textit{de novo} ligand design, integrating continuous and categorical flow matching with equivariant optimal transport and efficient protein pocket conditioning. Our empirical evaluations demonstrate that~\textsc{Flowr} substantially surpasses existing state-of-the-art diffusion- and flow-based methods across multiple critical metrics, including ligand validity, pose accuracy, interaction recovery, and inference speed. Specifically,~\textsc{Flowr} achieves up to 70-fold faster inference compared to diffusion-based methods.

Recognizing critical limitations in existing benchmark datasets, we also presented~\textsc{Spindr}, a rigorously curated dataset of ligand-pocket co-crystal complexes. By addressing prevalent structural defects, accurately inferring atomic-resolution interactions, protonation states and minimizing data leakage,~\textsc{Spindr} establishes a robust and realistic benchmark for evaluating generative models in structure-based drug discovery. Our results underscore the importance of high-quality datasets in accurately assessing model performance and generalization capabilities, and we encourage the adoption of~\textsc{Spindr} as a high-quality resource for future research.

Furthermore, we introduced~\textsc{Flowr.multi}, a versatile extension enabling targeted ligand generation conditioned on predefined interaction profiles or chemical substructures. Our experiments illustrate that~\textsc{Flowr.multi} substantially enhances interaction recovery and ligand quality, making it particularly suitable for fragment-based drug design tasks such as scaffold hopping, scaffold elaboration, and fragment-based ligand optimization. We demonstrated the practical utility of~\textsc{Flowr.multi} in hit expansion scenarios on two randomly selected test set targets, highlighting its ability to generate ligands closely aligned with desired interaction patterns and chemical constraints.

Collectively, our contributions represent advancements in AI-driven structure-based drug discovery. By combining state-of-the-art generative modeling techniques with rigorous dataset curation,~\textsc{Flowr} and~\textsc{Flowr.multi} provide powerful, efficient, and reliable tools for ligand generation. These models are applicable across diverse drug discovery scenarios, including hit identification, scaffold elaboration, and fragment-based design. We envision the \textsc{Spindr} dataset becoming a robust and challenging resource for evaluating future SBDD models.
Overall, we hope these developments will facilitate broader adoption of generative models in early-stage drug discovery.

\section*{Broader Impact}

\added{\textsc{Flowr} addresses a critical challenge in early drug discovery by enabling structure-aware ligand generation for hit-to-lead and lead optimization campaigns. As an ideation tool, it can accelerate the exploration of chemically diverse binding modes and allows scaffold hopping as well as fragment growing and replacement strategies while maintaining geometric compatibility with target binding sites. By reducing reliance on extensive virtual screening or de novo design cycles, this approach has the potential to streamline drug development timelines and reduce associated costs. However, while structure-based generative models have the potential to accelerate early-stage drug discovery by serving as ideation tools from hit exploration to lead optimisation, they must be viewed as complementary to---rather than replacements for---medicinal chemistry expertise and experimental validation. The generated molecules represent hypotheses that require rigorous downstream evaluation, including synthesis feasibility assessment, \textit{in vitro} activity profiling, and binding pose confirmation through co-crystallography or cryo-EM.}

\section*{Limitations}

\added{Several limitations of the current approach should be acknowledged. First, ligand validity degrades substantially when modelling explicit hydrogens, which we attribute to limited training data coverage and the absence of explicit hydrogen modelling on the protein side. Addressing this will likely require both larger datasets and joint protein-ligand hydrogen modelling during training. Second, while the \textsc{Spindr} dataset provides high-quality co-crystal structures with careful split design, it offers limited chemical and conformational space coverage compared to the full drug-like molecule landscape; models trained on it may not generalise well to underrepresented chemotypes or binding site topologies. Third, the model operates on static protein structures and does not account for conformational flexibility, induced fit effects, or allosteric modulations, which are often critical for accurate binding mode prediction in real drug discovery campaigns. Finally, while strain energies of generated ligands are substantially lower than those of competing approaches, they remain elevated relative to co-crystal reference structures, indicating that further improvements in conformational accuracy are needed.}

\section*{Future Work}

\added{Several future directions could address these limitations and expand the applicability of flow-based generative models for structure-based drug design. Scaling to larger and more chemically diverse training sets, potentially incorporating data from predicted protein-ligand complexes, could improve coverage of chemical space and enhance explicit hydrogen modelling. Incorporating protein pocket flexibility---for example, by conditioning on ensembles of conformations or integrating molecular dynamics snapshots---would better capture the dynamic nature of protein-ligand interactions. Direct integration of pharmacokinetic property predictors, synthetic accessibility constraints, and selectivity objectives into the generative process or as guidance signals during inference would enhance practical utility for medicinal chemistry programmes. Prospective experimental validation campaigns, in which generated ligands are synthesised and profiled against their intended targets, are essential to assess real-world performance and identify failure modes not captured by computational benchmarks. Additionally, systematic evaluation of \textsc{Flowr.multi} on further fragment-based design tasks, including fragment linking and fragment growing, would provide a more comprehensive assessment of its multi-purpose capabilities.}

\section*{Data availability}
The CrossDocked2020 dataset~\cite{dataset:crossdocked} with pre-computed data splits can be downloaded from GitHub at https://github.com/pengxingang/Pocket2Mol/tree/main/data.
The~\textsc{Spindr} dataset can be downloaded from Zenodo at https://zenodo.org/records/15257565~\cite{data:spindr}.

\section*{Code availability}
The source code used in this study is available under the MIT License on GitHub at https://github.com/jule-c/flowr and on Zenodo~\cite{code:flowr}. Model weights can be downloaded from Zenodo at https://zenodo.org/records/15257565~\cite{weights:spindr}.

\newpage
\bibliography{science}
\bibliographystyle{plainnat}

\FloatBarrier
\clearpage
\newpage

\setcounter{section}{0}
\setcounter{subsection}{0}
\setcounter{figure}{0}
\setcounter{table}{0}
\setcounter{equation}{0}

\renewcommand{\figurename}{Supplementary Figure}
\renewcommand{\tablename}{Supplementary Table}
\renewcommand{\thefigure}{S\arabic{figure}}
\renewcommand{\thetable}{S\arabic{table}}
\renewcommand{\theequation}{S\arabic{equation}}
\renewcommand{\thesection}{\arabic{section}}
\renewcommand{\thesubsection}{\arabic{section}.\arabic{subsection}}

\makeatletter
\renewcommand{\@seccntformat}[1]{%
  \ifnum\pdfstrcmp{#1}{section}=0
    Supplementary Section \csname the#1\endcsname:\quad
  \else
    \csname the#1\endcsname\quad
  \fi
}
\renewcommand{\l@section}[2]{%
  \begingroup
    \let\numberline\@gobble
    \@dottedtocline{1}{0em}{0em}{#1}{#2}%
  \endgroup
}
\renewcommand{\l@subsection}{\@dottedtocline{2}{1.5em}{2.3em}}
\makeatother

\begin{center}
{\Huge\bfseries Supplementary Information\par}
\vspace{0.8em}
{\large \scititle\par}
\vspace{2em}
\end{center}

\tableofcontents
\newpage

\section{Additional Details on the~\textsc{Spindr} Dataset}
\label{sec:spire-add}

\subsection{\textsc{Spindr} vs.~\textsc{CrossDocked}}

\begin{figure}[ht]
    \centering
    \includegraphics[width=1.0\columnwidth]{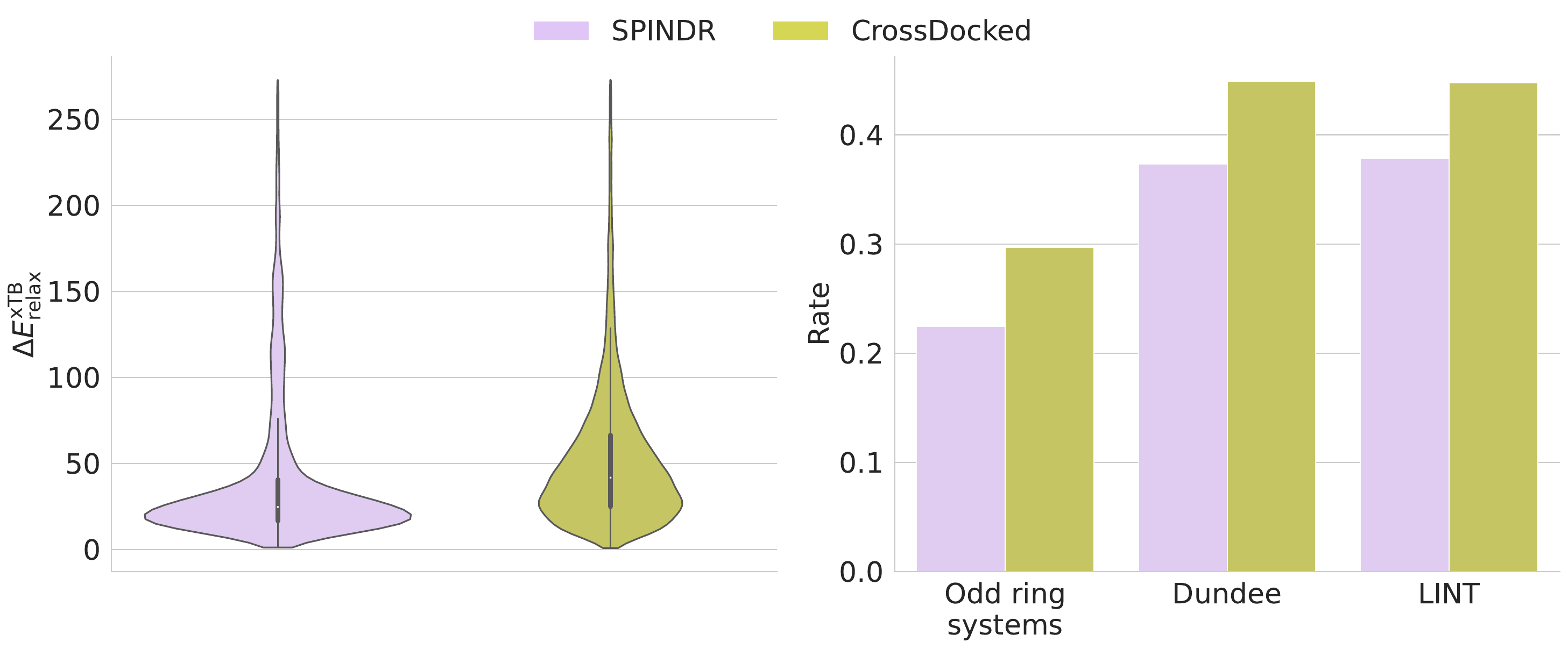}
    \caption{Comparison of the~\textsc{Spindr} and~\textsc{CrossDocked2020} datasets in terms of GFN2-xTB relaxation energies computed using the ALPB solvation model (left), as well as the proportion of unusual ring systems and potentially reactive functional groups identified according to two distinct substructure libraries, Dundee and LINT.}
    \label{fig:spindr_crossdocked_energy_walters}
\end{figure}

Supplementary Figure~\ref{fig:spindr_crossdocked_energy_walters} compares the quality of ligands in the~\textsc{Spindr} dataset against those in the~\textsc{CrossDocked2020} dataset, using three distinct metrics: relaxation energies using GFN2-xTB~\cite{xtb_gfn2} together with the ALPB solvation model~\cite{alpb}, frequency of odd ring systems, and presence of potentially reactive functional groups~\cite{walters_filter}. Lower delta relaxation energies indicate higher conformational quality as ligands require less structural rearrangement upon binding.
As illustrated, the~\textsc{Spindr} dataset exhibits substantially lower delta relaxation energies compared to~\textsc{CrossDocked2020}. Additionally, following~\cite{walters_filter}, odd ring systems are defined as ring structures that occur infrequently (fewer than 100 occurrences) within the ChEMBL database, indicating unusual or potentially problematic chemical motifs. Reactive functional groups were identified using standard medicinal chemistry filters, specifically the Dundee and LINT substructure collections. On all three metrics,~\textsc{Spindr} shows on average better values than the~\textsc{CrossDocked2020} dataset.

\added{Supplementary Table~\ref{tab:supp-ligand-comparison} further compares the ligand diversity of \textsc{Spindr} to the standard 100K \textsc{CrossDocked} benchmark dataset. While \textsc{Spindr} contains fewer systems than \textsc{CrossDocked}, it actually contains more unique ligands, and substantially more unique ligand scaffolds. Ultimately, this is due to \textsc{CrossDocked} being a synthetically-generated dataset created by cross-docking systems from PDBBind, which contains approximately 19K systems in total.}

\begin{table*}[ht]
    \small
    \caption{Ligand comparison in benchmark datasets. Scaffolds were obtained using RDKit's Murcko Scaffold utilities.}
    \label{tab:supp-ligand-comparison}
    \centering
    \added{
    \begin{tabular}{l|cc}
    \toprule
    Metric & \textsc{CrossDocked2020} & \textsc{Spindr} \\
    \midrule
    Total ligands              & 100,563    & 35,666 \\
    Unique ligands             & 8,433      & 11,689 \\
    Unique ligand scaffolds    & 4,855      & 7,713  \\
    \bottomrule
    \end{tabular}}
\end{table*}

\begin{table*}[ht]
    \small
    \caption{Sizes of train, validation and test dataset splits for the three proposed versions of the~\textsc{Spindr} dataset.}
    \label{tab:dedup-datasets}
    \centering
    \begin{tabular}{l|ccc}
    \toprule
    Dataset & Train Systems & Val Systems & Test Systems \\
    \midrule
   ~\textsc{Spindr}                             & 35,373  & 68    & 225 \\
   ~\textsc{Spindr}\textsuperscript{RMSD}       & 24,885  & 68    & 225  \\
   ~\textsc{Spindr}\textsuperscript{RMSD-SEQID} & 20,349  & 68    & 225  \\
    \bottomrule
    \end{tabular}
\end{table*}

Like existing datasets of protein-ligand complexes, the~\textsc{Spindr} training set contains many redundant systems -- systems which have substantial structural similarity to another training system. Understanding the impact of this redundancy on model performance is a relatively unexplored topic but could have an important influence on the design of future datasets. We therefore apply two data deduplication strategies to~\textsc{Spindr} and report results on all three datasets. Deduplication is only applied to the training data and all models are evaluated identically.

Our initial deduplication strategy works by creating groups of systems such that all systems within the group have identical ligands (based on their canonical SMILES after hydrogen atoms have been removed) and identical pockets atoms where the pocket coordinates are within an RMSD of 1.0 of some reference system for the group. We find that for system groups defined like this the distribution of RMSD values to the reference is very close to zero, so the choice of reference system and the RMSD threshold is not so important. In practice we iterate over all systems in the dataset, if a system cannot be added to an existing group a new group is created with this system as the group's reference system. Once all systems in the training dataset have been grouped a single system is randomly selected from each group to form the deduplicated training set. We refer to this dataset as~\textsc{Spindr}\textsuperscript{RMSD}. We also explore an extension of this deduplication strategy which allows systems to be in the same group if the sequence identity between a query pocket and the pocket of the reference system for a group is greater than 90\%. In this case the RMSD between the query and reference pockets is taken by comparing the coordinates only on matching residues. Again, once groups have been constructed, a single system is randomly sampled from each group to form the deduplicated training set. We refer to this dataset as~\textsc{Spindr}\textsuperscript{RMSD-SEQID}. The sizes of the three versions of the dataset are shown in Supplementary Table~\ref{tab:dedup-datasets}.

\FloatBarrier
\section{Additional Model Details}
\label{section:appendix-model}

Here we provide the methodological foundations and notation for the flow matching framework used in \textsc{Flowr}.
\added{Our approach builds on continuous normalizing flows (CNFs) trained via conditional flow matching (CFM)~\citep{cfm:general-dfm,cfm:gaussian-fm}, which combines the stable regression objectives of diffusion models with efficient deterministic inference in a simulation-free framework. We specifically employ optimal transport CFM (OT-CFM)~\citep{cfm:minibatch-ot} to construct simpler, more stable flows by minimising transport costs between source and target distributions. Extending this with equivariant flow matching~\citep{cfm:equi-ot} allows us to exploit the rotational and translational symmetries inherent to molecular systems, yielding flows with shorter integration paths, improved sampling efficiency, and natural incorporation of physical symmetries---essential for generating geometrically valid protein-ligand complexes. Below, we detail each of these components and describe how they are integrated into our model.}

\subsection{Flow Matching for Continuous Data}

Flow matching~\cite{cfm:gaussian-fm,cfm:si,cfm:reflow} is a generative modelling framework based on continuous normalizing flows (CNFs). A CNF defines a time-dependent flow $\phi_t: \mathbb{R}^d \to \mathbb{R}^d$, where $d$ is the data dimensionality, through the ordinary differential equation (ODE)
\begin{equation}
\frac{d}{dt}\phi_t(x) = v_t(\phi_t(x)), \quad \phi_0(x) = x,
\label{eq:cnf-ode}
\end{equation}
where $v_t: \mathbb{R}^d \times [0,1] \to \mathbb{R}^d$ is a time-dependent vector field. The flow $\phi_t$ pushes forward an initial (prior) distribution $p_0$ to a target (data) distribution $p_1$ through the induced time-dependent density path $p_t$. Samples from $p_1$ are obtained by drawing $x_0 \sim p_0$ and integrating the ODE forward from $t{=}0$ to $t{=}1$.

Training a CNF via maximum likelihood requires simulating the ODE trajectory and computing the divergence of $v_t$, which is computationally expensive. Flow matching provides a simulation-free regression objective by regressing a parameterised vector field $v_t^\theta(x_t)$, where $\theta$ denotes the learnable model parameters, against a target vector field $u_t(x_t)$:
\begin{equation}
\mathcal{L}_{\text{FM}}(\theta) = \mathbb{E}_{t \sim \mathcal{U}[0,1],\, x_t \sim p_t(x_t)} \|v_t^\theta(x_t) - u_t(x_t)\|^2.
\label{eq:fm-loss}
\end{equation}
In practice, both the marginal vector field $u_t(x_t)$ and the marginal probability path $p_t(x_t)$ are intractable. The crucial insight of conditional flow matching (CFM)~\cite{cfm:gaussian-fm,cfm:minibatch-ot} is that equivalent gradients can be obtained by conditioning on data samples $x_1 \sim p_1$. One defines a conditional probability path $p_{t|1}(x_t | x_1)$ with an associated conditional vector field $u_t(x_t | x_1)$, leading to the CFM objective:
\begin{equation}
\mathcal{L}_{\text{CFM}}(\theta) = \mathbb{E}_{t \sim \mathcal{U}[0,1],\, x_1 \sim p_1,\, x_t \sim p_{t|1}(x_t|x_1)} \|v_t^\theta(x_t) - u_t(x_t | x_1)\|^2.
\label{eq:cfm-loss}
\end{equation}
A common choice of conditional path is the Gaussian interpolation
\begin{equation}
p_{t|1}(x_t | x_1) = \mathcal{N}(x_t \mid t\, x_1 + (1 - t)\, x_0,\, \sigma^2 I),
\label{eq:gaussian-interp}
\end{equation}
where $x_0 \sim p_0$ is a sample from the prior distribution, $\sigma > 0$ is a small noise parameter, and $I$ denotes the identity matrix. The corresponding conditional vector field is $u_t(x_t | x_1) = x_1 - x_0$, defining straight conditional paths between noise and data that can be efficiently integrated during sampling.

\subsection{Mini-Batch Optimal Transport}

The choice of coupling between prior samples $x_0$ and data samples $x_1$ substantially affects the geometry of the learned flows. When $x_0$ and $x_1$ are coupled independently, that is, $(x_0, x_1) \sim p_0(x_0)\,p_1(x_1)$, the resulting marginal vector field can exhibit crossing paths and unnecessarily complex trajectories. Optimal transport CFM (OT-CFM)~\cite{cfm:minibatch-ot} addresses this by replacing the independent coupling with an approximate optimal transport plan $\pi(x_0, x_1)$ that minimises the expected squared Euclidean transport cost.

Computing the exact OT plan between continuous distributions is intractable in general. OT-CFM therefore approximates the OT coupling at the mini-batch level: given a batch of prior samples $(x_0^1, \dots, x_0^B)$ and data samples $(x_1^1, \dots, x_1^B)$, one computes the pairwise cost matrix $M_{ij} = \|x_0^i - x_1^j\|^2$ and solves the resulting discrete OT problem to obtain an optimal assignment within the batch. Training pairs are then formed according to this assignment. This mini-batch OT coupling produces straighter conditional paths, more stable training, and faster inference since the learned vector fields require fewer integration steps~\cite{cfm:minibatch-ot}.

\subsection{Equivariant Optimal Transport}

For molecular systems that exhibit symmetries under the Euclidean group $E(3)$ and the permutation group $S_N$, standard OT-CFM can yield suboptimal transport plans because it does not account for the invariance of the target distribution under rotations, reflections, and atom permutations. Equivariant optimal transport flow matching~\cite{cfm:equi-ot} addresses this by replacing the standard squared Euclidean cost with a symmetry-aware cost function:
\begin{equation}
\tilde{c}(x_0, x_1) = \min_{g \in G} \|x_0 - \rho(g)\, x_1\|^2,
\label{eq:equi-ot-cost}
\end{equation}
where $G$ is the relevant symmetry group and $\rho(g)$ denotes its action on the molecular configuration. For molecules, $G$ comprises rotations and reflections $O(D)$ combined with atom permutations $S(N)$, where $D{=}3$ is the spatial dimension and $N$ is the number of atoms. The cost function aligns each pair of samples along their symmetry orbits before computing the transport cost.

In practice, jointly optimising over all rotations and permutations is computationally intractable, so the minimisation is approximated sequentially~\cite{cfm:equi-ot}: first, the optimal permutation $\tilde{s} = \arg\min_{s \in S_N} \|x_0 - \rho(s)\, x_1\|^2$ is found using the Hungarian algorithm; then, the optimal rotation $R^* = \arg\min_{R \in O(D)} \|x_0 - R\,\rho(\tilde{s})\, x_1\|^2$ is computed via the Kabsch algorithm. The modified cost matrix is then used within the standard mini-batch OT solver. This equivariant OT procedure produces nearly optimal integration paths even for small batch sizes, leading to shorter paths, reduced integration errors, and improved sampling fidelity for molecular systems.

\subsection{Discrete Flow Models}

For categorical molecular features such as atom types and bond orders, we adopt the discrete flow model (DFM) framework~\cite{model:multiflow,cfm:discrete-fm}, which extends flow matching to discrete state spaces via continuous-time Markov chains (CTMCs). Analogously to continuous flow matching, DFM defines a conditional probability path $p_{t|1}(\cdot|x_1)$ that interpolates between a uniform prior and the data distribution. For a categorical variable $x_1$ with $K$ possible states, the conditional probability at time $t$ is given by
\begin{equation}
p_{t|1}(x_t | x_1) = t\, \delta(x_t, x_1) + (1-t)\, \frac{1}{K},
\label{eq:discrete-interp}
\end{equation}
where $\delta(x_t, x_1)$ is the Kronecker delta. At $t{=}0$ this reduces to a uniform distribution over all $K$ states, while at $t{=}1$ it concentrates on the data $x_1$. A neural network learns a data denoiser $p_{1|t}^{\theta}(\cdot | x_t)$ that predicts the clean data from the noisy sample, trained by minimising the cross-entropy between the predicted and true posterior distributions. During inference, the denoiser is used within a CTMC integration scheme that progressively drives $x_t$ from the uniform prior towards the data distribution, analogously to the Euler integration used for continuous variables.

\subsection{The \textsc{Semla} Architecture and \textsc{Flowr} Extensions}

The neural network architecture for \textsc{Flowr} is based on \textsc{Semla}~\cite{model:semlaflow}, a scalable E(3)-equivariant message passing architecture originally proposed in the \textsc{SemlaFlow} framework for unconditional 3D molecular generation. \textsc{Semla} represents each atom $i$ with invariant features $\mathbf{h}_i \in \mathbb{R}^{d_{\text{inv}}}$ and equivariant features $\mathbf{x}_i \in \mathbb{R}^{3 \times d_{\text{equi}}}$, where $d_{\text{inv}}$ and $d_{\text{equi}}$ denote the invariant and equivariant feature dimensions, respectively. Translation invariance is enforced through zero-centring of equivariant features; combined with equivariant updates throughout the network, the learned density is $E(3)$-invariant.

A key innovation in \textsc{Semla} is \emph{latent attention}: invariant node features are projected into a smaller latent space of dimension $d_l \ll d_{\text{inv}}$ before computing pairwise messages, reducing the computational complexity of the attention mechanism from $\mathcal{O}(N^2 d_{\text{inv}}^2)$ to $\mathcal{O}(N^2 d_l^2)$, where $N$ is the number of atoms. Pairwise messages are computed using a 2-layer MLP that combines latent invariant features with dot products of equivariant features, and are then split into separate invariant and equivariant attention scores. Softmax-normalised attention weights aggregate node features with a variance-preserving scheme.

We extend \textsc{Semla} for \textsc{Flowr} in several ways. First, we incorporate a separate pocket encoder that processes the protein pocket $\mathcal{P}$ independently of the flow time $t$ and the noisy ligand $l_t$. This encoder uses the same \textsc{Semla} layer design and outputs pocket embeddings that are reused across all integration steps during ligand generation, substantially reducing computational cost. Second, we add a cross-attention module within the ligand decoder that takes invariant and equivariant embeddings of $\mathcal{P}$, $l_t$, and optionally the interaction matrix $\mathcal{I}$, following the same latent attention design for efficiency.

Third, we replace the equivariant feed-forward module in \textsc{Semla} with a gated variant. For atom $i$ with invariant features $\mathbf{h}_i \in \mathbb{R}^{d_{\text{inv}}}$ and equivariant features $\mathbf{x}_i \in \mathbb{R}^{3 \times d_{\text{equi}}}$, the gated feed-forward output is:
\begin{equation}
\mathbf{x}_i^{\text{out}} = \mathbf{W}_{\theta}^2 \hspace{2pt} \mathbf{\hat{x}}_i, \quad \text{where} \quad \mathbf{\hat{x}}_i = \sigma ( \Phi_{\theta} (\mathbf{h}_i, \|\mathbf{x}_i\|) ) \hspace{2pt} \odot \hspace{2pt} \mathbf{W}_{\theta}^1 \hspace{2pt} \mathbf{x}_i,
\label{eq:gated-ff}
\end{equation}
where $\sigma$ denotes the elementwise sigmoid function, $\odot$ is elementwise multiplication, $\mathbf{W}_{\theta}^1, \mathbf{W}_{\theta}^2 \in \mathbb{R}^{d_{\text{equi}} \times d_{\text{equi}}}$ are learnable weight matrices, and $\Phi_{\theta}$ is a two-layer MLP mapping invariant features and equivariant norms to gating coefficients. This gated module is substantially faster than the original equivariant feed-forward block. Fourth, we pass bond embeddings into the self-attention module on every layer, as opposed to only the first layer, improving molecular validity with negligible impact on inference time.

We parameterise \textsc{Flowr} with a 4-layer pocket encoder using $d_{\text{inv}}^{\text{enc}} = 256$ and a 12-layer ligand decoder using $d_{\text{inv}}^{\text{dec}} = 384$. Both encoder and decoder use $d_{\text{equi}} = 64$, a latent attention dimension of $d_l = 64$, and $n_{\text{heads}} = 32$ attention heads.

\subsection{Training and Inference}

We train \textsc{Flowr} to generate ligands conditioned on a given structure. Since 3D molecular graphs contain a mixture of continuous and categorical data types, \textsc{Flowr} jointly generates continuous and discrete distributions. Our approach follows a similar setup to~\cite{model:semlaflow}. Specifically, we apply the continuous flow matching framework from~\cite{cfm:minibatch-ot} to learn ligand coordinates, and the discrete flow models framework from~\cite{model:multiflow} to learn atom types and bond orders. Ligand formal charges are not learned through a flow, but simply predicted by the model.

\subsection{Model Training}

Training proceeds by sampling ligand noise $l_0 \sim p_{\text{noise}}$, a ligand, pocket and interaction tuple $(l_1, \mathcal{P}, \mathcal{I}) \sim p_{\text{data}}$, and a time $t \in [0,1]$. We use Gaussian noise for coordinates and uniform distributions for atom and bond types to create $p_{\text{noise}}$. Writing $l_t = (\mathbf{x}_t, \mathbf{a}_t, \mathbf{b}_t)$ for the coordinate, atom type, and bond order components of the noisy ligand, we sample from the conditional probability path $l_t \sim p_{t|1}(l_t|l_1)$ used in~\cite{model:semlaflow}, defined as:
\begin{gather}
    t \sim \text{Beta} (\alpha, \beta)
    \hspace{20pt}
    \mathbf{x}_t \sim \mathcal{N}(t \mathbf{x}_1 + (1-t) \mathbf{x}_0, \sigma^2)
    \\
    \mathbf{a}_t \sim \text{Cat}(t \delta(\mathbf{a}_1) + (1-t) \frac{1}{|A|})
    \hspace{15pt}
    \mathbf{b}_t \sim \text{Cat}(t \delta(\mathbf{b}_1) + (1-t) \frac{1}{|B|})
\end{gather}
Here $\text{Cat}(\cdot)$ denotes the categorical distribution, $A$ and $B$ are the sets of possible values for atom types and bond orders, respectively, and $\delta(\cdot)$ is the one-hot encoding operation applied to each item in a sequence individually. We use values $\alpha = 2.0$, $\beta = 1.0$, and $\sigma = 0.2$ for all \textsc{Flowr} models.

Following~\cite{model:midi,model:eqgatdiff,model:pilot} we train \textsc{Flowr} to predict $l_1$ directly by learning the distribution $p^{\theta}_{1|t} (l_1 | l_t, \mathcal{P}, \mathcal{I})$. This leads to the same loss function as \textsc{SemlaFlow}~\cite{model:semlaflow} -- we apply a mean-squared error loss for ligand coordinates and cross-entropy losses for atom types, bond orders and formal charges. When the model is conditioned on both $\mathcal{P}$ and $\mathcal{I}$, the interaction features are provided as additional input to the cross-attention module, enabling the model to attend to both structural pocket information and desired interaction patterns simultaneously. In Section~\ref{section:appendix-interactions} we provide more information on how we handle the case where the model is conditioned on both $\mathcal{P}$ and $\mathcal{I}$.

Additionally, during training we apply self-conditioning~\cite{self-cond} to improve generation quality. In self-conditioned training, half of the training batches are processed normally, while the other half first generate a preliminary prediction of $l_1$ from the model, which is then detached from the computation graph and provided as additional conditioning input in the subsequent forward pass. For atom and bond types, the conditioning inputs are softmax-normalised probability distributions over the predicted categorical types.

We also apply equivariant optimal transport~\cite{cfm:equi-ot} during training to reduce the transport cost between $p_{\text{noise}}$ and $p_{\text{data}}$. For each training pair $(l_0, l_1)$, the coordinate noise $\mathbf{x}_0$ is transformed via $\hat{\mathbf{x}}_0 = f_\pi(\mathbf{x}_0, \mathbf{x}_1)$, where $f_\pi$ applies the permutation and rotation that minimise the squared error between $\mathbf{x}_0$ and $\mathbf{x}_1$, as described in the Equivariant Optimal Transport section above. This alignment reduces transport costs and yields straighter, more efficient integration paths during sampling.

\added{\subsection{Extracting Fragments}

We extract molecular scaffolds using RDKit's \texttt{GetScaffoldForMol} from the \texttt{MurckoScaffold} implementation, defining functional groups as all atoms not part of the scaffold and linkers as non-ring scaffold atoms. To enable diverse fragment-based learning, we additionally employ RDKit's matched molecular pairs analysis (rdMMPA) via \texttt{FragmentMol} to decompose molecules into chemically meaningful fragments, randomly sampling from these at each training batch. This strategy allows the model to learn scaffold hopping and fragment growing patterns naturally from the data. For interaction-conditional training, we leverage \texttt{ProLIF}-derived interaction fingerprints pre-calculated for all complexes in the \textsc{Spindr} dataset.}

\subsection{Generating Ligands}

Given a protein pocket $\mathcal{P}$ and, optionally, a desired interaction matrix $\mathcal{I}$, we can generate samples from the learned data distribution by setting $l_t \leftarrow l_0$ where $l_0 \sim p_{\text{noise}}$ and pushing $l_t$ toward the data distribution by following the learned vector field. Specifically, for molecular coordinates $\mathbf{x}_t$ we follow the vector field $v_t^{\theta}(\mathbf{x}_t) = \frac{1}{1-t} (\tilde{\mathbf{x}}_1 - \mathbf{x}_t)$ where $\tilde{\mathbf{x}}_1$ is the coordinate component of $\tilde{l}_1 \sim p_{1|t}^{\theta}(l_1|l_t, \mathcal{P}, \mathcal{I})$. We then integrate the vector field using an Euler solver with step size $\Delta t$ as follows: $\tilde{\mathbf{x}}_{t+\Delta t} = \mathbf{x}_t + \Delta t \hspace{3pt} v_t^{\theta}(\mathbf{x}_t)$. For discrete atom and bond types, integration proceeds analogously: at each step, the model predicts the posterior distribution $p_{1|t}^{\theta}(x_1 | x_t)$ over categorical types, and $x_t$ is updated towards the data distribution according to the CTMC integration scheme, where the transition rates are derived from the predicted posteriors and the conditional probability path defined in the Discrete Flow Models section.

\paragraph{Evaluation}

To maintain consistency across models, we used identical random seeds for training, inference, and data loading. Additionally, we applied the same sampling and evaluation scripts across all models. For each of the 225 test set targets, we generated 100 ligand samples using a standardized size sampling approach. Specifically, we determined native ligand sizes and applied a uniform sampling scheme, allowing for a size deviation of -25\% to +10\%. This procedure was performed using the same seed across all models to ensure direct comparability.

\FloatBarrier
\section{Additional Experimental Results}
\label{sec:appendix-results}

Benchmarking newly proposed models and architectures in the context of structure-based drug design requires careful consideration of multiple evaluation aspects. In addition to the results presented in the main text, we provide a broader assessment using various metrics and evaluation settings in the following sections.
Specifically, we evaluate the novelty of generated ligands with respect to the training set, as well as the average uniqueness and diversity among the 100 generated ligands per target. To ensure a comprehensive analysis, we consider both SMILES string- and ECFP4-based measures for uniqueness and diversity. Additionally, following \cite{benchmarking:genbench3d}, we extend this analysis to include conformer-based uniqueness and diversity.
As indicators of drug-likeness, we report RDKit's Quantitative Estimate of Drug-likeness (QED), the Synthetic Accessibility Score (SAScore)~\cite{benchmarking:sa}, molecular weight, logP values, and compliance with Lipinski's Rule of Five.

\FloatBarrier
\subsection{\textsc{Flowr} vs.~\textsc{Pilot}}
\label{app:full_eval}

\begin{table*}[t!]
\caption{Benchmark of the proposed~\textsc{Flowr} model against the recent state-of-the-art diffusion-based~\textsc{Pilot} model on the~\textsc{Spindr} dataset. We report RDKit- and PoseBusters-validity of generated ligands, the GenBench3D strain energy and the AutoDock-Vina score. We also state the Wasserstein distance of generated ligands for the bond angles and bond lengths distribution to the~\textsc{Spindr} test set. Novelty, uniqueness and diversity measure the capability of the model to explore the chemical space both in 2D and 3D. RDKit's QED evaluation, SAScore, the molecular weight as well as the logP values evaluate drug-likeness of generated ligands. All presented values are mean values taken for 100 sampled ligands per test set target. The test dataset comprises 225 test set targets. Ligand sizes were drawn from a uniform distribution around the ground truth ligand size allowing for a deviation of -10\% and + 10\% with the same random seed for all models. Note, both RDKit- and PoseBusters-validity are evaluated on the raw generated set of 100 ligands per target. All other metrics are calculated on the subset of RDKit-valid ligands.}
\label{tab:full_results}
\centering
\begin{adjustbox}{width=1.0\textwidth,center}
\begin{sc}
\begin{tabular}{l|c|cc|cc}
\toprule
Metric & Test set &~\textsc{Pilot}$^{\text{no-Hs}}$ &~\textsc{Pilot}$^{\text{with-Hs}}$ &~\textsc{Flowr}$^{\text{no-Hs}}$ &~\textsc{Flowr}$^{\text{with-Hs}}$ \\
\midrule
RDKit-validity & 1.00 {\tiny$\pm$ 0.00}  & 0.79 {\tiny$\pm$ 0.39} & 0.52 {\tiny$\pm$ 0.50} & 0.94 {\tiny$\pm$ 0.24} & 0.64 {\tiny$\pm$ 0.48} \\
PB-validity & 0.99 {\tiny$\pm$ 0.02}  & 0.71 {\tiny$\pm$ 0.18} & 0.47 {\tiny$\pm$ 0.14} & 0.88 {\tiny$\pm$ 0.21} & 0.60 {\tiny$\pm$ 0.22} \\
\midrule
Strain energy & 43.27 {\tiny$\pm$ 41.85} & 120.10 {\tiny$\pm$ 71.61} & 53.07 {\tiny$\pm$ 22.84} & 90.05 {\tiny$\pm$ 52.18} & 54.11 {\tiny$\pm$ 33.36} \\
Vina score & -7.69 {\tiny$\pm$ 2.00} & -6.30 {\tiny$\pm$ 0.96} & -5.00 {\tiny$\pm$ 0.65} & -6.93 {\tiny$\pm$ 0.92} & -6.48 {\tiny$\pm$ 0.87} \\
Vina score (minimized) & -7.88 {\tiny$\pm$ 2.00} & -6.68 {\tiny$\pm$ 1.07} & -5.50 {\tiny$\pm$ 0.66} & -7.22 {\tiny$\pm$ 0.92} & -6.86 {\tiny$\pm$ 0.87} \\
\midrule
BondAnglesW1 & - & 1.82 & 2.81 & 1.08 & 0.82 \\
BondLengthsW1 [10$^{-2}$] & - & 0.42 & 0.10 & 0.35 & 0.11\\
\midrule
Novelty & 1.00 {\tiny$\pm$ 0.00} & 0.99 {\tiny$\pm$ 0.10} & 1.00 {\tiny$\pm$ 0.00} & 0.94 {\tiny$\pm$ 0.23} & 1.00 {\tiny$\pm$ 0.00} \\
Uniqueness2D & 0.92 {\tiny$\pm$ 0.10} & 0.99 {\tiny$\pm$ 0.05} & 1.00 {\tiny$\pm$ 0.02} & 0.94 {\tiny$\pm$ 0.13} & 0.97 {\tiny$\pm$ 0.07} \\
Uniqueness3D & - & 0.66 {\tiny$\pm$ 0.20} & 0.59 {\tiny$\pm$ 0.19} & 0.50 {\tiny$\pm$ 0.20} & 0.55 {\tiny$\pm$ 0.17} \\
Diversity2D & 0.92 {\tiny$\pm$ 0.04} & 0.89 {\tiny$\pm$ 0.03} & 0.90 {\tiny$\pm$ 0.02} & 0.86 {\tiny$\pm$ 0.05} & 0.87 {\tiny$\pm$ 0.06} \\
Diversity3D & - & 0.25 {\tiny$\pm$ 0.13} & 0.13 {\tiny$\pm$ 0.19} & 0.21 {\tiny$\pm$ 0.12} & 0.18 {\tiny$\pm$ 0.11} \\
\midrule
SA & 0.66 {\tiny$\pm$ 0.12} & 0.63 {\tiny$\pm$ 0.12} & 0.64 {\tiny$\pm$ 0.10} & 0.67 {\tiny$\pm$ 0.13} & 0.65 {\tiny$\pm$ 0.10} \\
QED & 0.49 {\tiny$\pm$ 0.22} & 0.51 {\tiny$\pm$ 0.21} & 0.53 {\tiny$\pm$ 0.18} & 0.52 {\tiny$\pm$ 0.21} & 0.53 {\tiny$\pm$ 0.21} \\
Rings & 2.98 {\tiny$\pm$ 1.42} & 2.52 {\tiny$\pm$ 1.42} & 1.52 {\tiny$\pm$ 0.98} & 2.68 {\tiny$\pm$ 1.35} & 2.64 {\tiny$\pm$ 1.43} \\
Aromatic Rings & 1.84 {\tiny$\pm$ 1.31} & 1.12 {\tiny$\pm$ 1.07} & 1.21 {\tiny$\pm$ 0.95} & 1.52 {\tiny$\pm$ 1.16} & 1.59 {\tiny$\pm$ 1.22} \\
HAcceptors & 7.30 {\tiny$\pm$ 4.49} & 6.19 {\tiny$\pm$ 3.30} & 5.46 {\tiny$\pm$ 2.21} & 6.67 {\tiny$\pm$ 4.23} & 6.47 {\tiny$\pm$ 3.64} \\
HDonors & 2.62 {\tiny$\pm$ 1.68} & 2.52 {\tiny$\pm$ 1.65} & 1.55 {\tiny$\pm$ 1.27} & 2.52 {\tiny$\pm$ 1.68} & 2.66 {\tiny$\pm$ 1.58} \\
LogP & 0.29 {\tiny$\pm$ 3.48} & 0.45 {\tiny$\pm$ 3.08} & -0.03 {\tiny$\pm$ 2.33} & 0.29 {\tiny$\pm$ 3.31} & 0.34 {\tiny$\pm$ 2.99} \\
MolWt & 390.43 {\tiny$\pm$ 119.82} & 336.79 {\tiny$\pm$ 107.86} & 337.30 {\tiny$\pm$ 83.59} & 350.10 {\tiny$\pm$ 114.00} & 336.09 {\tiny$\pm$ 108.60} \\
Lipinski & 4.00 {\tiny$\pm$ 1.34} & 4.45 {\tiny$\pm$ 0.93} & 4.73 {\tiny$\pm$ 0.55} & 4.35 {\tiny$\pm$ 1.05} & 4.32 {\tiny$\pm$ 1.05} \\
\bottomrule
\end{tabular}
\end{sc}
\end{adjustbox}
\end{table*}

\begin{figure}[t!]
    \centering
    \includegraphics[width=1.0\textwidth]{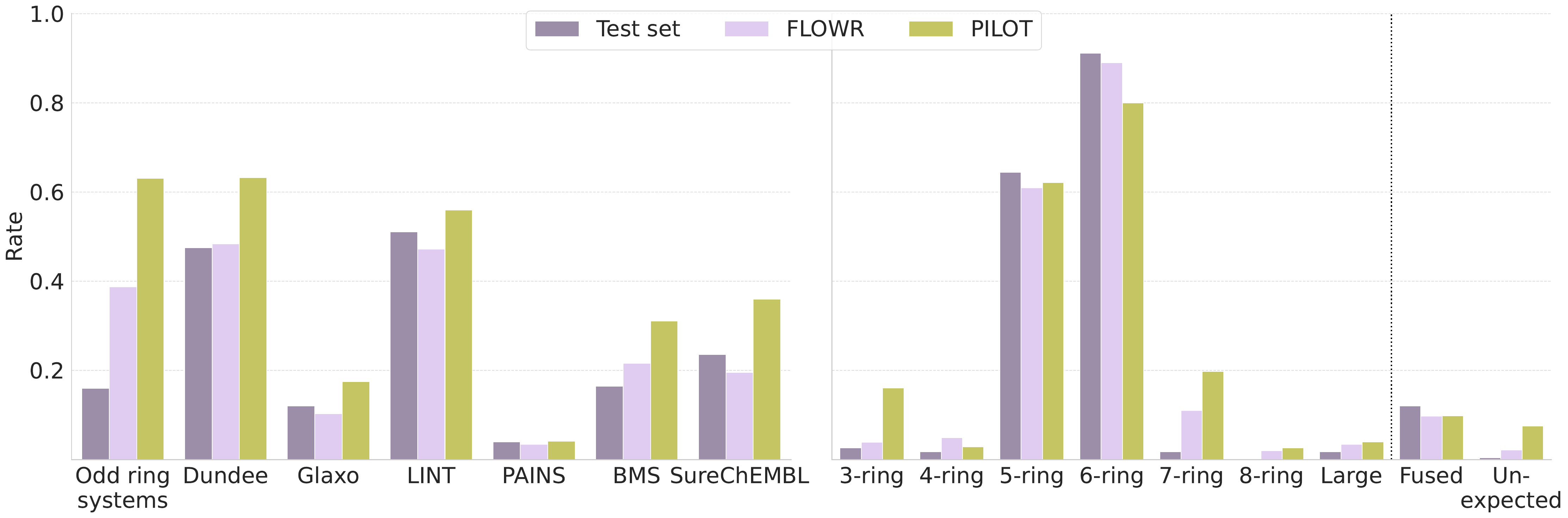}
    \caption{\textbf{Comparison of~\textsc{Flowr} and~\textsc{Pilot} using the 'Walters'-filter and ring distribution analysis.} We assess the distribution learning capabilities of~\textsc{Flowr} and compare its performance against~\textsc{Pilot}, utilizing established medicinal chemistry filters provided by the \href{https://github.com/PatWalters/useful_rdkit_utils}{\textsc{useful\_rdkit\_utils}} toolkit (left panel). Specifically, we employed the~\textsc{REOS} filter~\cite{walters_filter} and the~\textsc{RingSystemLookup} to evaluate the generated ligands.  Additionally, we present a comparative analysis of ring distributions (right panel). All reported metrics represent mean values computed from 100 ligands sampled per target using the~\textsc{Spindr} test dataset, which comprises 225 distinct test set targets. Ligand sizes were sampled uniformly around the ground truth ligand size, allowing for deviations within a range of $\pm$10\%, using a consistent random seed across models.}
    \label{fig:flowr_pilot_walters}
\end{figure}

In Supplementary Table~\ref{tab:full_results}, we report the results comparing~\textsc{Pilot} and~\textsc{Flowr} for both settings, without explicit and with explicit hydrogens in training and inference, respectively. On average,~\textsc{Pilot} shows higher novelty, uniqueness and diversity values of generated ligands. However, in light of the substantially worse results across distribution and ligand-pocket-centric metrics, it is likely that~\textsc{Pilot} has a stronger tendency to hallucinate and thus generates physically less plausible, but more diverse structures with higher strains.
Regarding RDKit-based ligand property metrics, both models show similar results, while~\textsc{Flowr} shows in general a higher overlap with the test set values indicating better distribution learning capabilities.

Additionally, we compare~\textsc{Flowr} and~\textsc{Pilot} in terms of drug-likeness filtering using the 'Walters'-filter~\cite{walters_filter}, which evaluates generated compounds for unusual ring systems (by comparing their frequencies against ring systems found in ChEMBL) and identifies problematic functional groups through substructure matching against established filter collections, including Dundee, Glaxo, LINT, PAINS, BMS, and SureChEMBL. Supplementary Figure~\ref{fig:flowr_pilot_walters} (left) illustrates the performance comparison between~\textsc{Flowr} and~\textsc{Pilot} based on these drug-likeness criteria. We observe that~\textsc{Flowr} consistently outperforms~\textsc{Pilot}, in some cases by substantial margins, and generates compounds whose properties align more closely with those observed in the test set. Furthermore, Supplementary Figure~\ref{fig:flowr_pilot_walters} (right) presents a comparative analysis of ring distributions, demonstrating that~\textsc{Flowr} again achieves substantially better overlap with the~\textsc{Spindr} test set distribution compared to~\textsc{Pilot}.

\FloatBarrier
\subsection{Performance of \textsc{Flowr.multi} on~\textsc{Spindr}}
\label{app:flowr.multi_eval}
Here we report the evaluation results for~\textsc{Flowr.multi} using different conditional generation settings on the~\textsc{Spindr} test dataset. Supplementary Table~\ref{tab:flowr.multi-results} shows an overview of a set of evaluation metrics to assess pose quality and distribution learning capabilities.

\begin{table}[t!]
\caption{We evaluate~\textsc{Flowr.multi} for different conditional modes, namely for interaction-, scaffold-, functional-group- and linker-conditional generation, on the~\textsc{Spindr} test set. We report RDKit- and PoseBusters-validity of generated ligands, the GenBench3D strain energy and the AutoDock-Vina score. We also state the pocket-ligand interaction fingerprint recovery rate and Tanimoto similarity using~\textsc{ProLIF}, and the Wasserstein distance of generated ligands for the bond angles and bond lengths distribution to the~\textsc{Spindr} test set. Novelty, uniqueness and diversity measure the capability of the model to explore the chemical space both in 2D and 3D. RDKit's QED evaluation, SAScore, the molecular weight as well as the logP values evaluate drug-likeness of generated ligands. All presented values are mean values taken for 100 sampled ligands per test set target. The test dataset comprises 225 test set targets. Ligand sizes were taken from the respective reference ligand and are not sampled. Note, both RDKit- and PoseBusters-validity values are evaluated on the generated set of 100 ligands per target. All other metrics are calculated on the subset of RDKit-valid ligands.}
\label{tab:flowr.multi-results}
\centering
\begin{adjustbox}{width=1.0\textwidth,center}
\begin{sc}
\begin{tabular}{lcccc}
\toprule
Metric & \textsc{Flowr.multi}$^{\mathrm{interact.-cond.}}$ & \textsc{Flowr.multi}$^{\mathrm{scaffold-cond.}}$ & \textsc{Flowr.multi}$^{\mathrm{f.-group-cond.}}$ & \textsc{Flowr.multi}$^{\mathrm{linker-cond.}}$ \\
\midrule
RDKit-validity         & 0.93\,$_{\scriptscriptstyle\pm0.25}$  & 0.93\,$_{\scriptscriptstyle\pm0.25}$  & 0.92\,$_{\scriptscriptstyle\pm0.26}$  & 0.92\,$_{\scriptscriptstyle\pm0.25}$  \\
PB-validity         & 0.86\,$_{\scriptscriptstyle\pm0.19}$  & 0.88\,$_{\scriptscriptstyle\pm0.13}$  & 0.86\,$_{\scriptscriptstyle\pm0.17}$  & 0.85\,$_{\scriptscriptstyle\pm0.17}$  \\
\midrule
Vina score          & -7.18\,$_{\scriptscriptstyle\pm0.83}$  & -7.41\,$_{\scriptscriptstyle\pm0.67}$  & -7.10\,$_{\scriptscriptstyle\pm0.71}$  & -7.35\,$_{\scriptscriptstyle\pm0.56}$  \\
Vina score (minimized)       & -7.48\,$_{\scriptscriptstyle\pm0.80}$  & -7.72\,$_{\scriptscriptstyle\pm0.59}$  & -7.34\,$_{\scriptscriptstyle\pm0.72}$  & -7.57\,$_{\scriptscriptstyle\pm0.56}$  \\
Strain energy   & 107.60\,$_{\scriptscriptstyle\pm93.07}$ & 86.26\,$_{\scriptscriptstyle\pm78.31}$  & 105.32\,$_{\scriptscriptstyle\pm95.47}$ & 94.56\,$_{\scriptscriptstyle\pm84.08}$  \\
\midrule
PLIF recovery rate  & 0.75\,$_{\scriptscriptstyle\pm0.08}$  & 0.65\,$_{\scriptscriptstyle\pm0.11}$  & 0.79\,$_{\scriptscriptstyle\pm0.12}$  & 0.79\,$_{\scriptscriptstyle\pm0.08}$  \\
PLIF Tanimoto similarity & 0.66\,$_{\scriptscriptstyle\pm0.09}$  & 0.62\,$_{\scriptscriptstyle\pm0.10}$  & 0.74\,$_{\scriptscriptstyle\pm0.13}$  & 0.76\,$_{\scriptscriptstyle\pm0.09}$  \\
\midrule
BondAnglesW1            & 1.17    & 0.84    & 1.14    & 0.91    \\
BondLengthsW1 [10$^{-2}$]     & 0.43  & 0.52  & 0.58  & 0.69  \\
\midrule
Novelty             & 0.93\,$_{\scriptscriptstyle\pm0.26}$  & 0.94\,$_{\scriptscriptstyle\pm0.23}$  & 0.87\,$_{\scriptscriptstyle\pm0.33}$  & 0.87\,$_{\scriptscriptstyle\pm0.33}$  \\
Uniqueness2D    & 0.83\,$_{\scriptscriptstyle\pm0.26}$  & 0.74\,$_{\scriptscriptstyle\pm0.28}$  & 0.70\,$_{\scriptscriptstyle\pm0.33}$  & 0.53\,$_{\scriptscriptstyle\pm0.31}$  \\
Uniqueness3D    & 0.40\,$_{\scriptscriptstyle\pm0.21}$  & 0.35\,$_{\scriptscriptstyle\pm0.12}$  & 0.31\,$_{\scriptscriptstyle\pm0.20}$  & 0.26\,$_{\scriptscriptstyle\pm0.17}$  \\
Diversity2D     & 0.82\,$_{\scriptscriptstyle\pm0.08}$  & 0.77\,$_{\scriptscriptstyle\pm0.07}$  & 0.78\,$_{\scriptscriptstyle\pm0.08}$  & 0.75\,$_{\scriptscriptstyle\pm0.06}$  \\
Diversity3D     & 0.06\,$_{\scriptscriptstyle\pm0.07}$  & 0.02\,$_{\scriptscriptstyle\pm0.01}$  & 0.07\,$_{\scriptscriptstyle\pm0.12}$  & 0.03\,$_{\scriptscriptstyle\pm0.05}$  \\
\bottomrule
\end{tabular}
\end{sc}
\end{adjustbox}
\end{table}

\FloatBarrier
\subsection{Strain Analysis}

\begin{figure}[t!]
    \centering
    \includegraphics[width=1.0\textwidth]{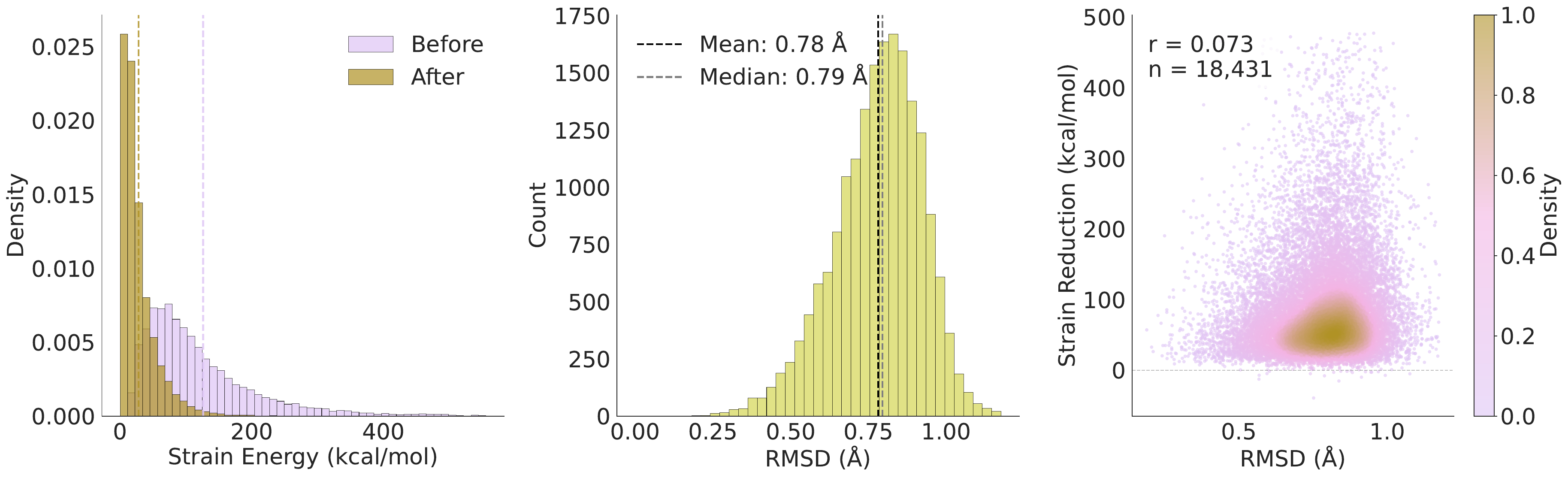}
    \caption{\added{\textbf{Strain analysis on the \textsc{Spindr} test set.} We run MMFF94s-based relaxation using RDKit on all generated ligands across the \textsc{Spindr} test set targets with protein pockets being fixed, and show pre- and post-relaxation strain energies and RMSDs.}}
    \label{fig:strain_analysis}
\end{figure}

To assess the physical realism of generated conformations, we performed energy minimization on all predicted poses. The optimization substantially reduced molecular strain energy, corresponding to a mean reduction of 62.75 kcal/mol. Importantly, this strain relief was achieved with minimal structural perturbation, as evidenced by the low RMSD of 0.78 $\pm$ 0.14 \AA{} between pre- and post-minimization conformations (see Supplementary Figure~\ref{fig:strain_analysis}). The refined poses exhibited improved molecular quality metrics, with PoseBusters validity increasing to 0.95 $\pm$ 0.08 and Vina score improving to -6.97 $\pm$ 0.89 kcal/mol. These results demonstrate that while the model generates physically plausible binding modes, local energy minimization can effectively relieve residual strain without fundamentally altering the predicted protein-ligand interactions.

\FloatBarrier
\subsection{Interactions}
\label{section:appendix-interactions}

\begin{figure}[t!]
    \centering
    \includegraphics[width=0.9\columnwidth]{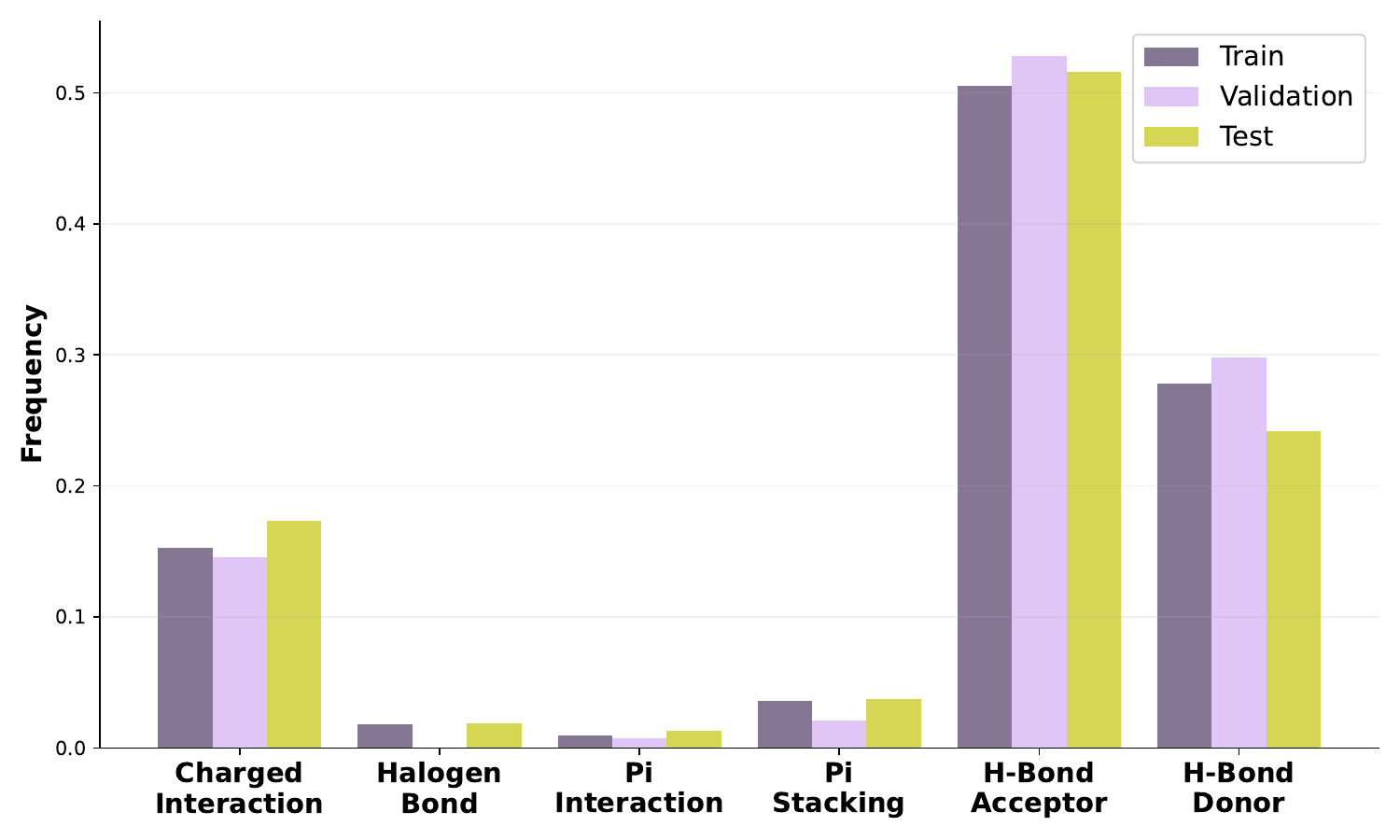}
    \caption{
    Distribution of interaction types on the train, validation and test sets of the~\textsc{Spindr} data that we considered in this work. Charged Interactions refers to either anionic or cationic interactions, and Pi Interactions refer to either cation-pi or pi-cation interactions. The \textsc{Spindr} dataset contains all possible interactions supported by ProLIF, but we focus on the above interactions for conditional generation since they are the most applicable for small molecule binding.}
    \label{fig:data-interactions}
\end{figure}

Following~\cite{bouysset_recovery_2024}, we consider a subset of interaction types in this work extracted using ProLIF~\cite{bouysset_prolif_2021}, including H-bonds (ligand acceptor and ligand donor), $\pi$-$\pi$ stacking, halogen bonds (ligand donor), $\pi$-cation (ligand $\pi$ / protein +), cation-$\pi$ (ligand + / protein $\pi$), anionic (ligand - / protein +), and cationic (ligand + / protein -) interactions. The distribution of these interactions within the~\textsc{Spindr} dataset is shown in Supplementary Figure~\ref{fig:data-interactions}. Notably, interaction sparsity is high, with an average of 99.85\% of ligand-protein atom pairs exhibiting no interactions.

\FloatBarrier
\subsection{Interactions Per Target}

\begin{figure}[t!]
    \centering
    \includegraphics[width=0.9\columnwidth]{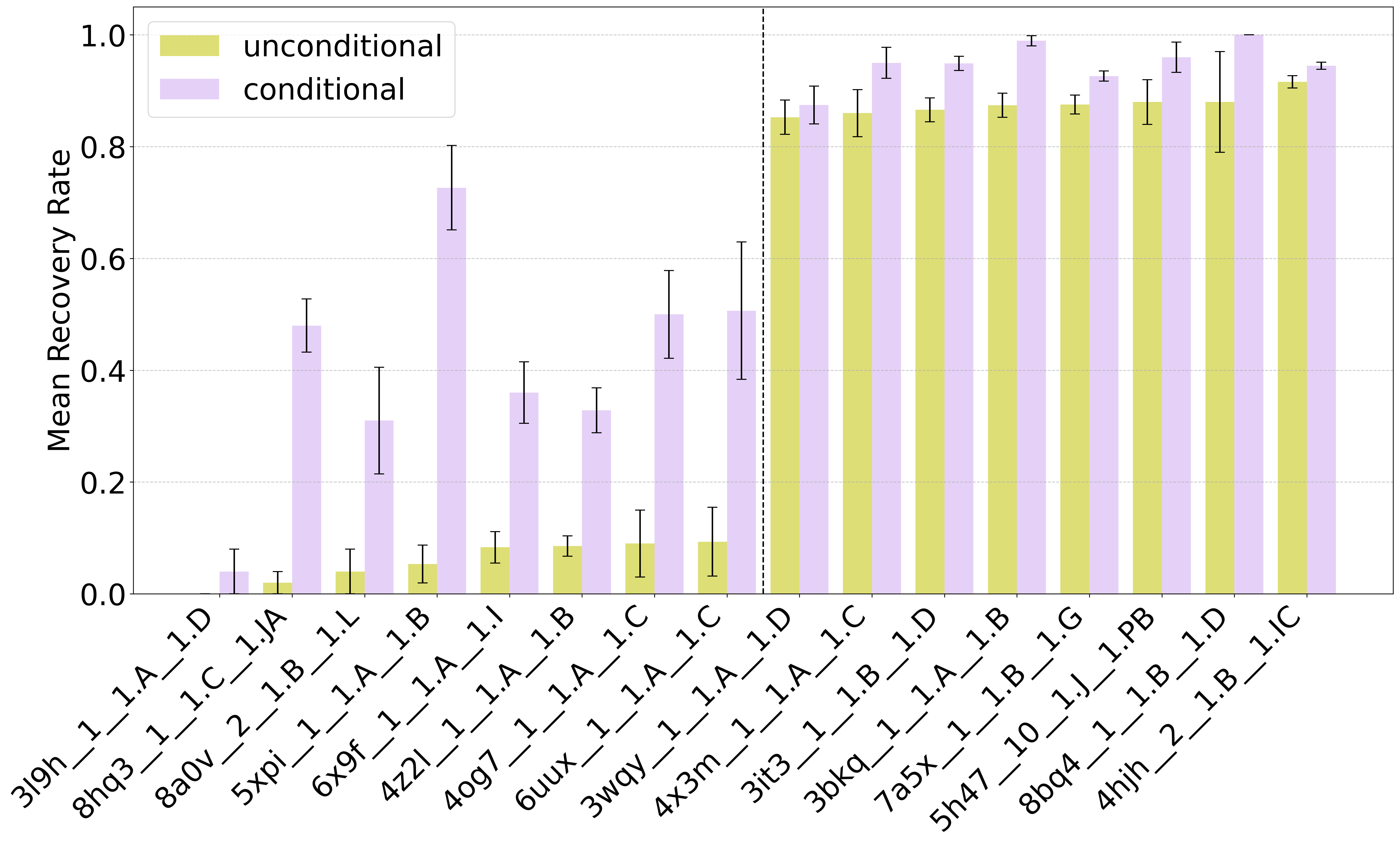}
    \caption{
    Comparison between~\textsc{Flowr} and~\textsc{Flowr.multi}. We identify eight targets with the lowest (left) and highest (right) average interaction recovery rates under the unconditional~\textsc{Flowr} model. For these selected targets, we compare the performance of the~\textsc{Flowr.multi} model to assess the impact of conditioning on pocket-ligand interactions.}
    \label{fig:interaction_recovery_per_target}
\end{figure}

To better evaluate the effectiveness of the proposed interaction-conditional training and sampling, we compare~\textsc{Flowr} with~\textsc{Flowr.multi} models on a per-target basis. Given that the test set comprises 225 targets, visualizing results for all targets is impractical. Instead, we select eight targets with the lowest and with the highest mean interaction recovery rates, as determined by the unconditional model, and compare the corresponding results obtained using the conditional model. This comparison is presented in Supplementary Figure~\ref{fig:interaction_recovery_per_target}. Notably, the conditional model consistently improves interaction recovery across targets where the unconditional model struggled to generate ligands with meaningful interactions. Additionally, it achieves substantially better results even for the top-performing targets, demonstrating that interaction-conditional generation effectively enhances ligand design with pre-specified interaction patterns.

\begin{figure}[t!]
  \centering
  \begin{tabular}{c|c}
    \subcaptionbox{Reference}[0.45\linewidth]{%
      \fbox{\includegraphics[width=\linewidth]{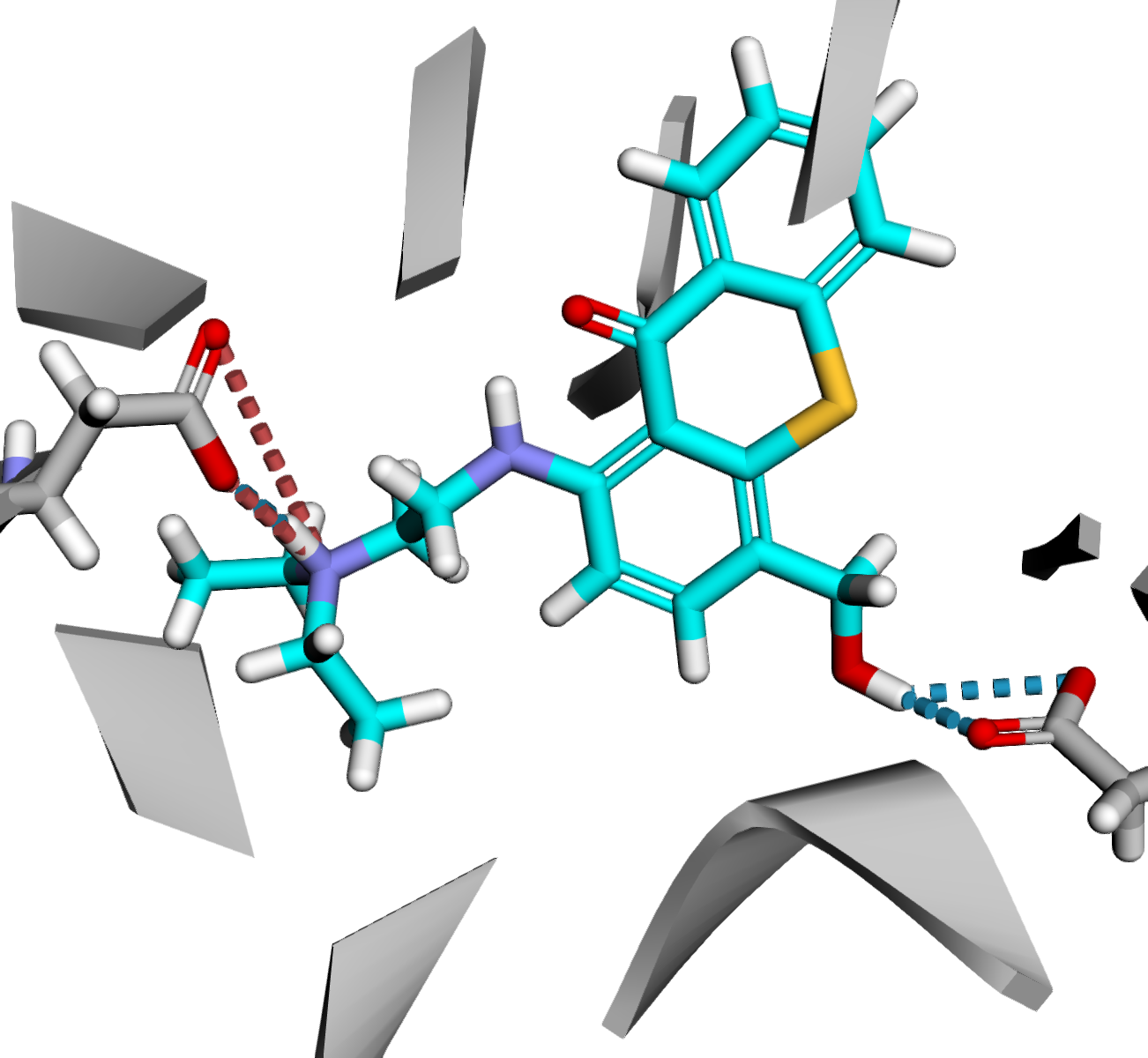}}} &
    \subcaptionbox{Prediction}[0.45\linewidth]{%
      \includegraphics[width=\linewidth]{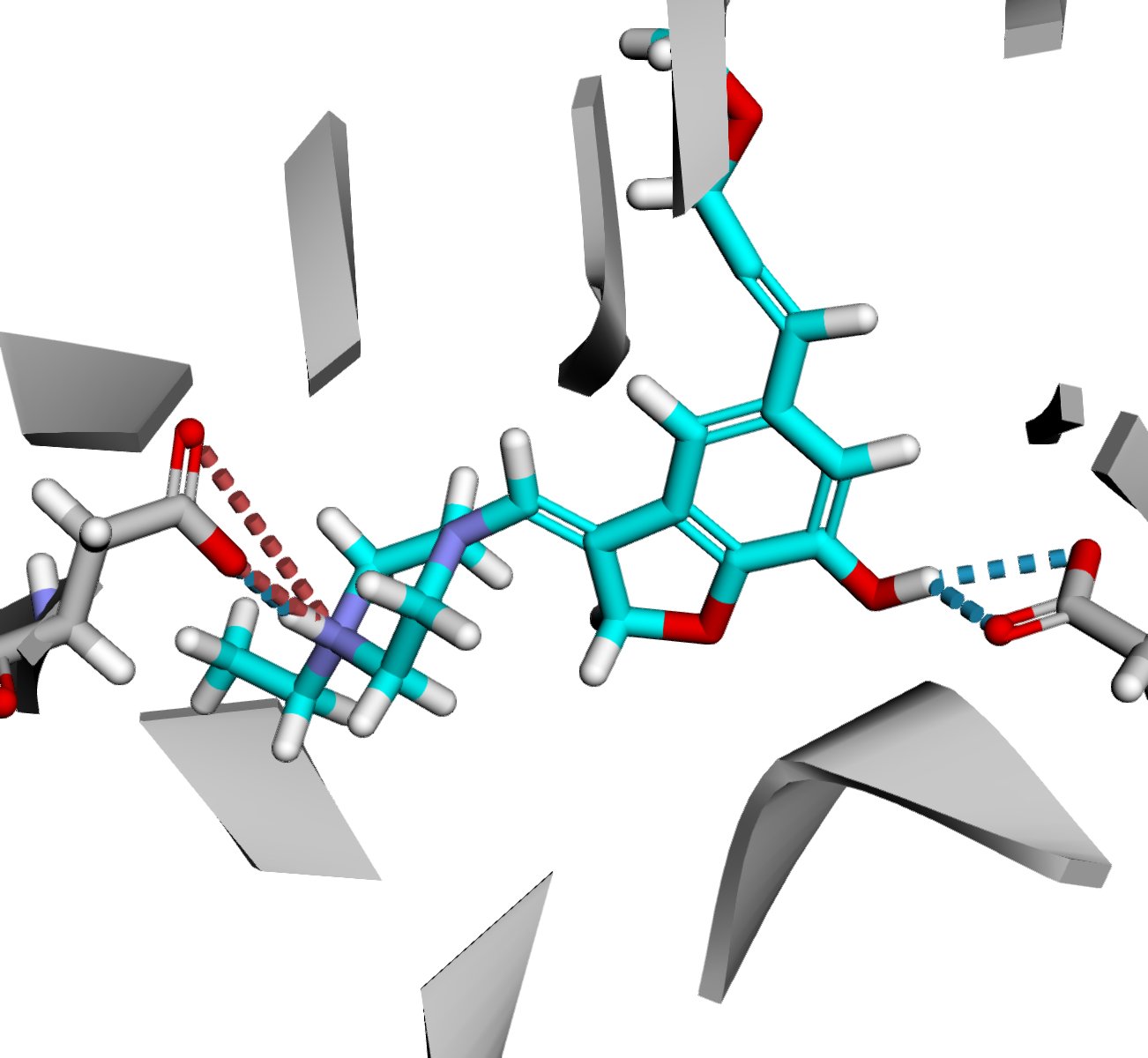}} \\
    \hline
    \subcaptionbox{Prediction}[0.45\linewidth]{%
      \includegraphics[width=\linewidth]{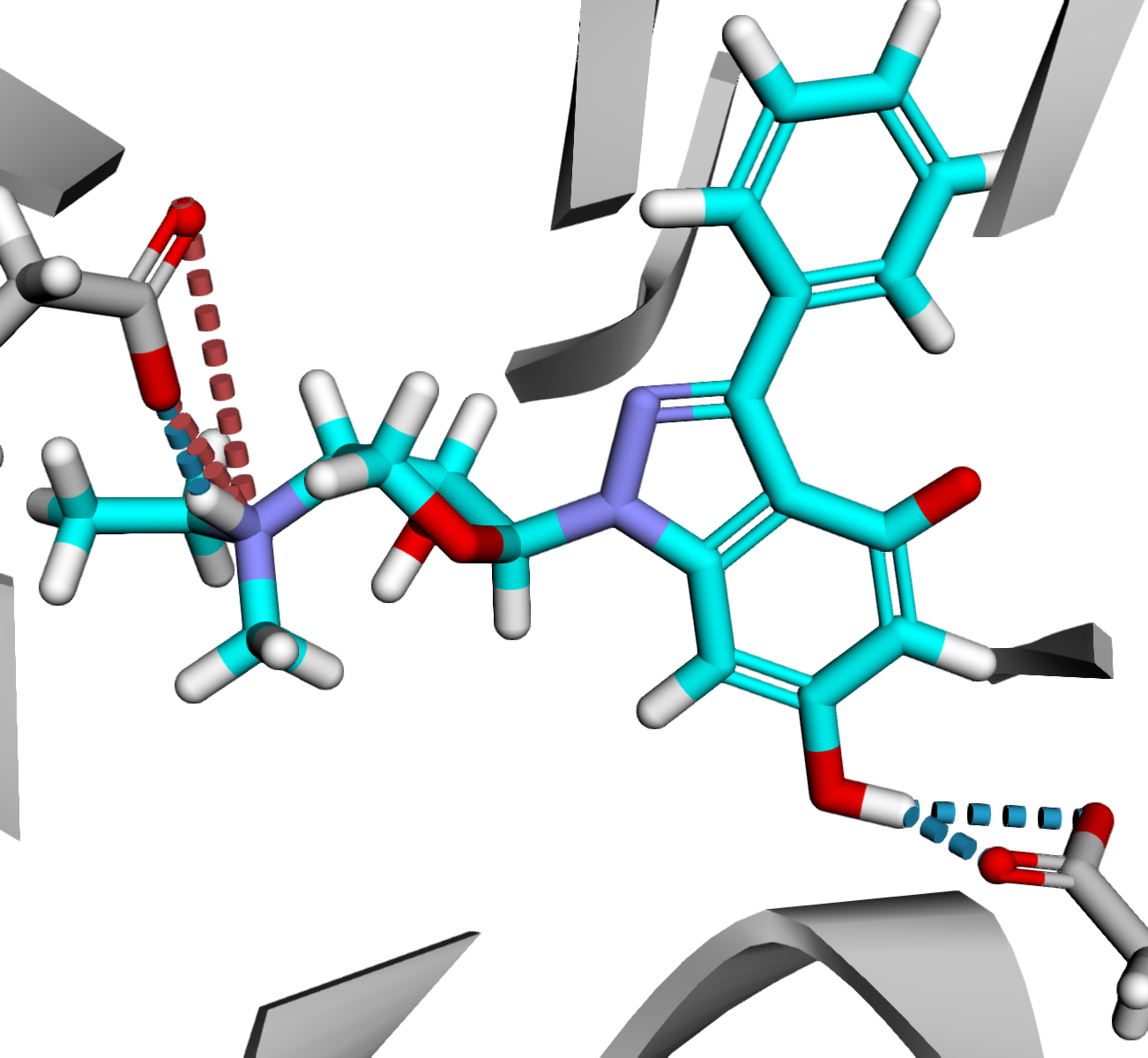}} &
    \subcaptionbox{Prediction}[0.45\linewidth]{%
      \includegraphics[width=\linewidth]{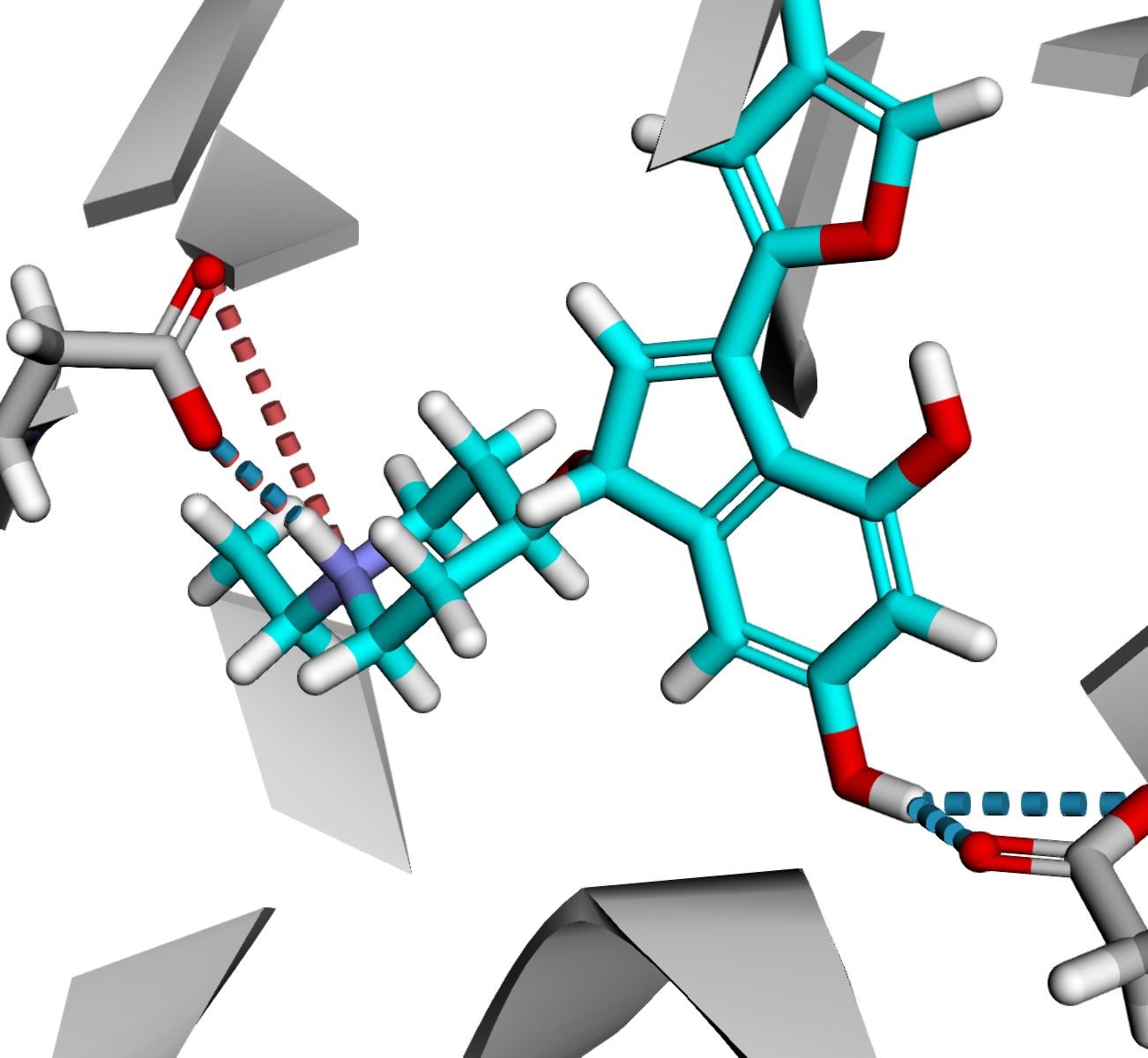}} \\
  \end{tabular}
  \caption{Comparison of reference and predicted ligands on their interaction profiles for the pocket of the protein with PDB id 6UUX sampled with~\textsc{Flowr.multi}. Atom colors: C (cyan/gray), N (blue), O (red), S (yellow), F (ochre), Cl (green), H (white)}
  \label{fig:comparison-interaction}
\end{figure}

Supplementary Figure~\ref{fig:comparison-interaction} presents an example of interaction profiling using the reference ligand of protein 6UUX alongside three randomly selected ligands generated by the interaction-conditional mode of~\textsc{Flowr.multi} model. The reference ligand forms two cationic interactions and one H-bond (ligand donor) interaction with ASP149, as well as two H-bond (ligand donor) interactions with ASP93. Notably, all of these interactions are successfully recovered in the generated ligands.

\FloatBarrier
\subsection{4MPE: Visualizations}

\begin{table}[t!]
    \caption{\textbf{Evaluation of~\textsc{Flowr.multi} on 4MPE.}
    Performance evaluation for interaction-, scaffold-, and functional group-conditional generation with~\textsc{Flowr.multi} on the test target with PDB-ID 4MPE. We report PoseBusters-validity (PB-validity) across 100 ligands per target, the mean Vina score (kcal/mol) as well as interaction recovery rate (PLIF recovery) and synthesizability score (SA score). }
    \label{tab:case_study_4mpe}
    \centering
    \begin{adjustbox}{width=1.0\textwidth,center}
    \begin{sc}
    \begin{tabular}{l|l|c|ccc}
        \toprule
        Protein & Metric & Reference &~\textsc{Flowr.multi}$^{\mathrm{interact.\text{-}cond.}}$ &~\textsc{Flowr.multi}$^{\mathrm{scaffold\text{-}cond.}}$ &~\textsc{Flowr.multi}$^{\mathrm{f.\,group-\,cond.}}$ \\
        \midrule
        \multirow{4}{*}{4MPE} & PB-validity $\uparrow$  & 1.0 & 0.95 & 1.0 & 0.92 \\[0.5ex]
                             & Vina score $\downarrow$    & -7.23 & -6.80 & -7.27 & -6.41 \\[0.5ex]
                             & Vina score (Top-10) $\downarrow$ & - & -7.54 & -7.83 & -7.15 \\[0.5ex]
                             & PLIF recovery rate $\uparrow$ & - & 0.79 & 0.53 & 0.89 \\[0.5ex]
                             & SA score $\uparrow$      & 0.84 & 0.81 & 0.82 & 0.82 \\
        \bottomrule
    \end{tabular}
    \end{sc}
    \end{adjustbox}
\end{table}

\begin{figure}[t!]
    \centering
    \includegraphics[width=1.0\textwidth]{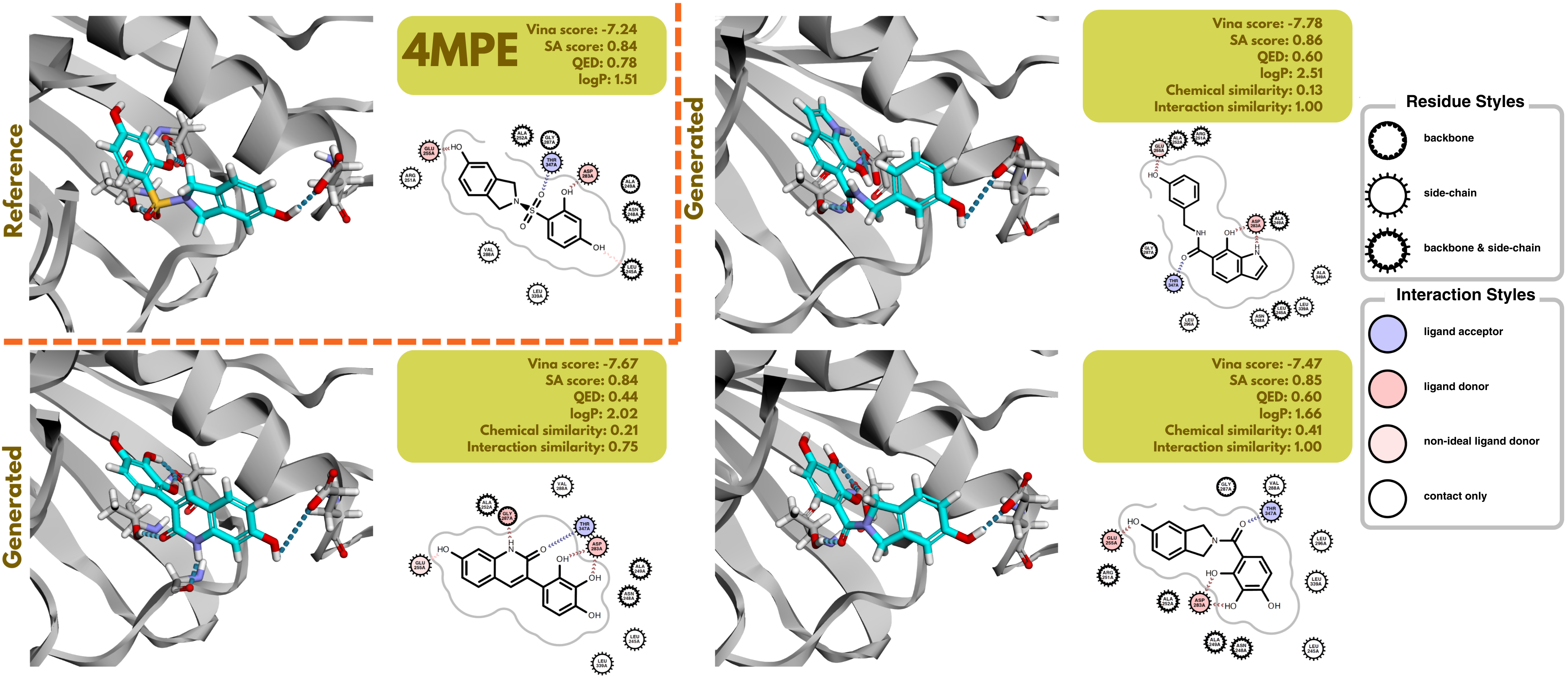}
    \caption{\textbf{Evaluation of interaction-conditional generation on 4MPE with~\textsc{Flowr.multi}}
    Using the interaction-conditional generation mode of~\textsc{Flowr.multi}, we sample 100 ligands for a randomly selected target from the~\textsc{Spindr} test set, here 4MPE. We then select three ligands at random and compare them to the reference compound based on Vina score, SA score, QED, logP, chemical similarity, and interaction similarity. Atom colors: C (cyan/gray), N (blue), O (red), S (yellow), F (ochre), Cl (green), H (white)}
    \label{fig:flowr_interact_cond_4mpe}
\end{figure}

\begin{figure}[t!]
    \centering
    \includegraphics[width=1.0\textwidth]{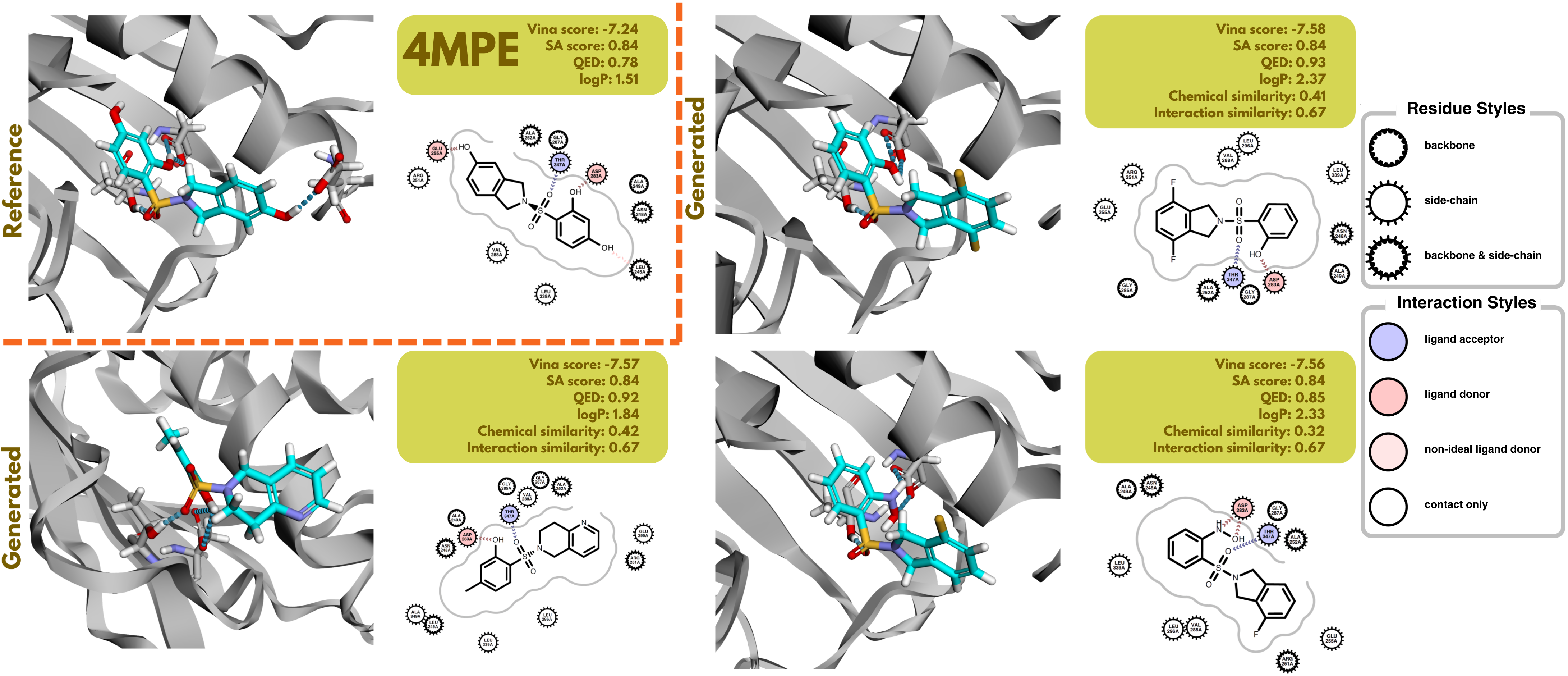}
    \caption{\textbf{Evaluation of scaffold-conditional generation on 4MPE with~\textsc{Flowr.multi}}
    Using the scaffold-conditional generation mode of~\textsc{Flowr.multi}, we sample 100 ligands for a randomly selected target from the~\textsc{Spindr} test set, here 4MPE. We then select three ligands at random and compare them to the reference compound based on Vina score, SA score, QED, logP, chemical similarity, and interaction similarity. Atom colors: C (cyan/gray), N (blue), O (red), S (yellow), F (ochre), Cl (green), H (white)}
    \label{fig:flowr_scaffold_cond_4mpe}
\end{figure}

\begin{figure}[t!]
    \centering
    \includegraphics[width=1.0\textwidth]{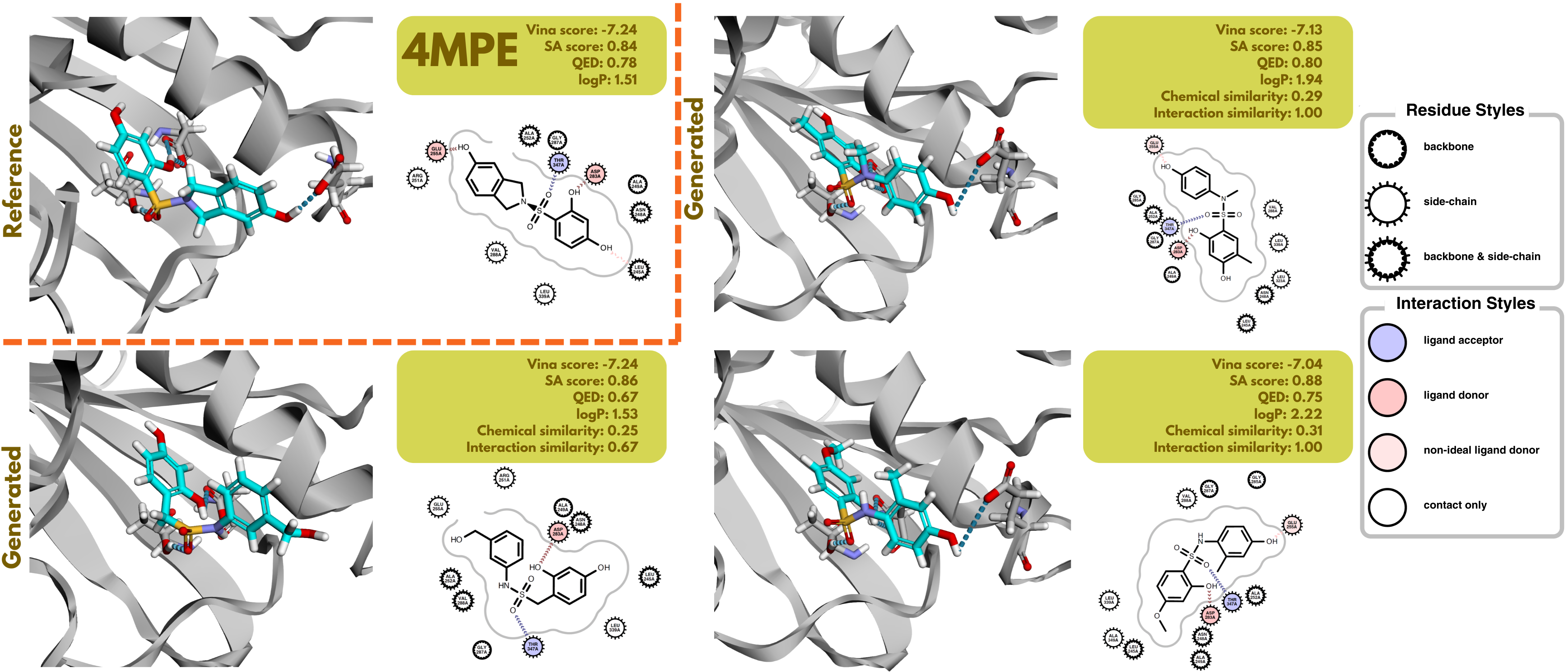}
    \caption{\textbf{Evaluation of functional-group-conditional generation on 4MPE with~\textsc{Flowr.multi}}
    Using the functional-group-conditional generation mode of~\textsc{Flowr.multi}, we sample 100 ligands for a randomly selected target from the~\textsc{Spindr} test set, here 4MPE. We then select three ligands at random and compare them to the reference compound based on Vina score, SA score, QED, logP, chemical similarity, and interaction similarity. Atom colors: C (cyan/gray), N (blue), O (red), S (yellow), F (ochre), Cl (green), H (white)}
    \label{fig:flowr_func_group_cond_4mpe}
\end{figure}

Here we show additional results on the protein target with PDB ID 4MPE for different conditional modes using the~\textsc{Flowr.multi} model.

\FloatBarrier
\subsection{5YEA: Visualizations}

\begin{table}[t!]
    \caption{\textbf{Evaluation of~\textsc{Flowr.multi} on 5YEA.}
    Performance evaluation for interaction-, scaffold-, and functional group-conditional generation with~\textsc{Flowr.multi} on a randomly selected test target with PDB-ID 5YEA. We report PoseBusters-validity (PB-validity) across 100 ligands per target, the mean Vina score (kcal/mol) as well as interaction recovery rate (PLIF recovery) and synthesizability score (SA score). }
    \label{tab:case_study_5yea}
    \centering
    \begin{adjustbox}{width=1.0\textwidth,center}
    \begin{sc}
    \begin{tabular}{l|l|c|ccc}
        \toprule
        Protein & Metric & Reference &~\textsc{Flowr.multi}$^{\mathrm{interact.\text{-}cond.}}$ &~\textsc{Flowr.multi}$^{\mathrm{scaffold\text{-}cond.}}$ &~\textsc{Flowr.multi}$^{\mathrm{f.\,group-\,cond.}}$ \\
        \midrule
        \multirow{4}{*}{5YEA} & PB-validity $\uparrow$  & 1.0 & 0.90  & 0.98  & 0.89  \\[0.5ex]
                             & Vina score $\downarrow$    & -9.57 & -8.96  & -8.71  & -8.99  \\[0.5ex]
                             & Vina score (Top-10) $\downarrow$ & - & -10.08 & -10.16 & -8.99 \\[0.5ex]
                             & PLIF recovery rate $\uparrow$ & - & 0.87  & 0.75  & 0.77  \\[0.5ex]
                             & SA score $\uparrow$      & 0.82 & 0.77 & 0.82 & 0.76 \\
        \bottomrule
    \end{tabular}
    \end{sc}
    \end{adjustbox}
\end{table}

\begin{figure}[t!]
    \centering
    \includegraphics[width=1.0\textwidth]{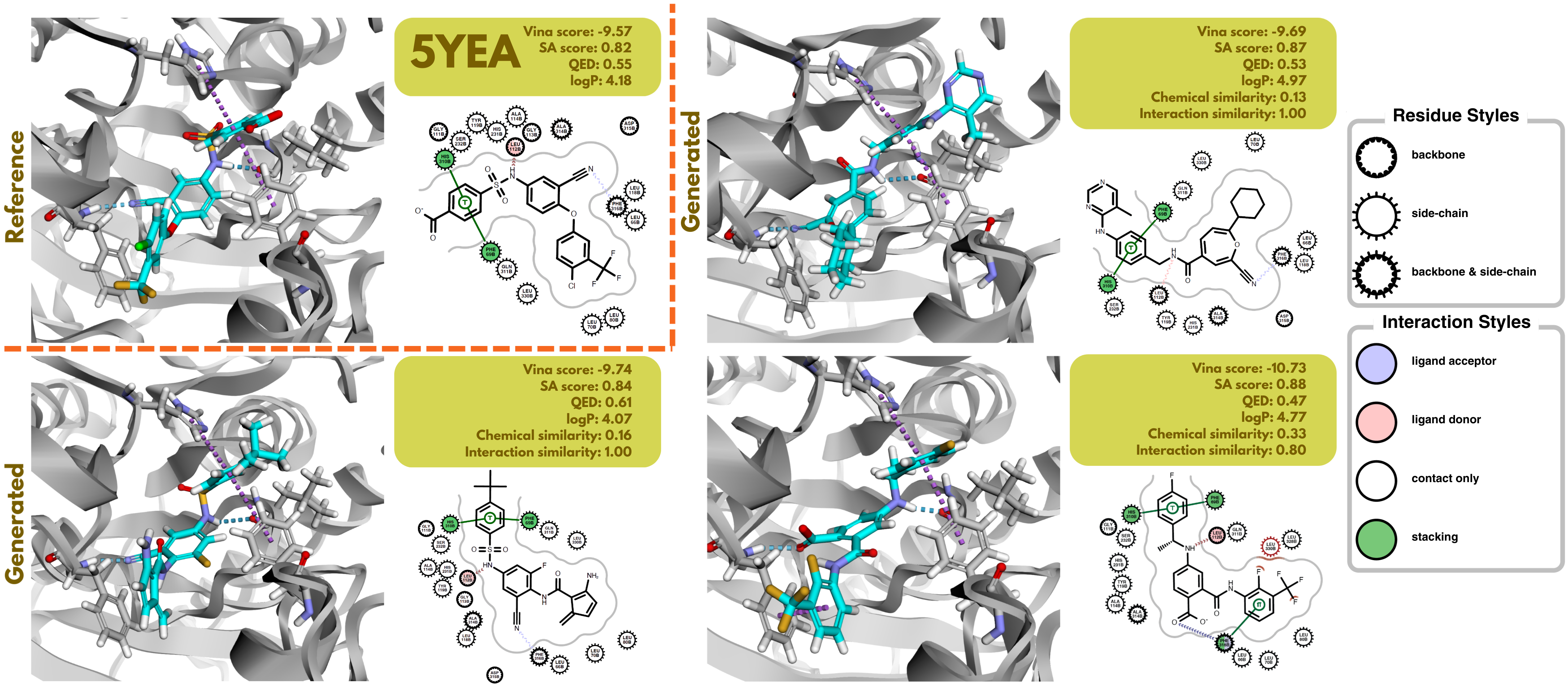}
    \caption{\textbf{Evaluation of interaction-conditional generation on 5YEA with~\textsc{Flowr.multi}}
    Using the interaction-conditional generation mode of~\textsc{Flowr.multi}, we sample 100 ligands for a randomly selected target from the~\textsc{Spindr} test set, here 5YEA. We then select three ligands at random and compare them to the reference compound based on Vina score, SA score, QED, logP, chemical similarity, and interaction similarity.}
    \label{fig:flowr_interact_cond_5yea}
\end{figure}

\begin{figure}[t!]
    \centering
    \includegraphics[width=1.0\textwidth]{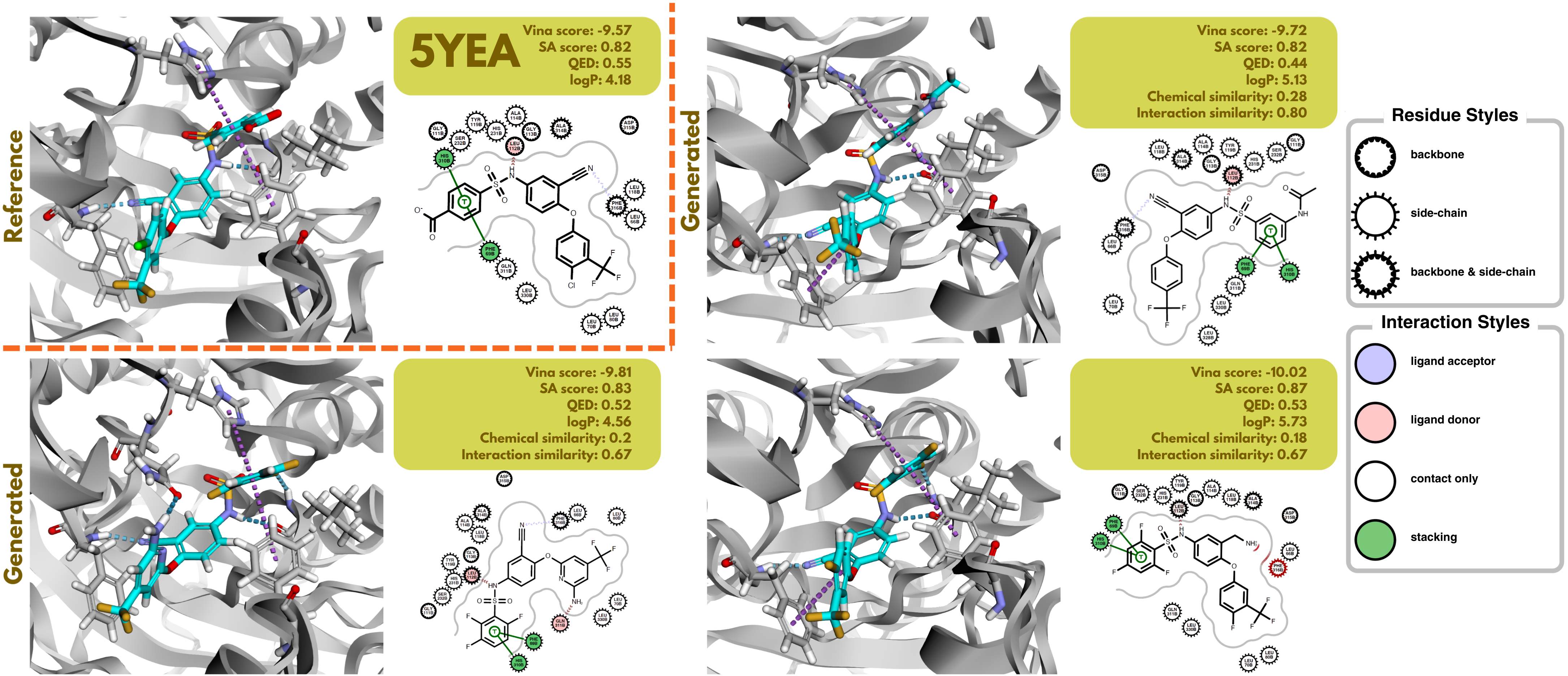}
    \caption{\textbf{Evaluation of scaffold-conditional generation on 5YEA with~\textsc{Flowr.multi}}
    Using the scaffold-conditional generation mode of~\textsc{Flowr.multi}, we sample 100 ligands for a randomly selected target from the~\textsc{Spindr} test set, here 5YEA. We then select three ligands at random and compare them to the reference compound based on Vina score, SA score, QED, logP, chemical similarity, and interaction similarity. Atom colors: C (cyan/gray), N (blue), O (red), S (yellow), F (ochre), Cl (green), H (white)}
    \label{fig:flowr_scaffold_cond_5yea}
\end{figure}

\begin{figure}[t!]
    \centering
    \includegraphics[width=1.0\textwidth]{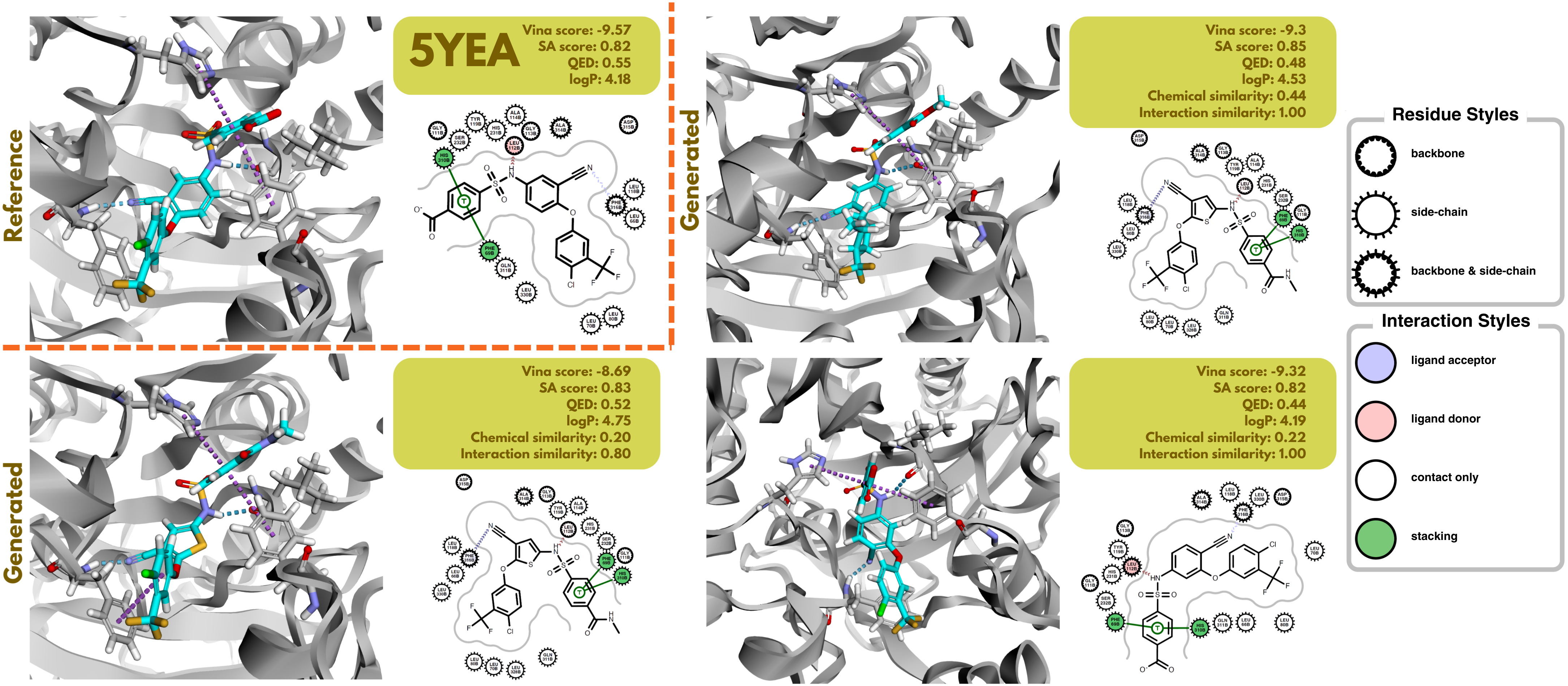}
    \caption{\textbf{Evaluation of functional-group-conditional generation on 5YEA with~\textsc{Flowr.multi}}
    Using the functional-group-conditional generation mode of~\textsc{Flowr.multi}, we sample 100 ligands for a randomly selected target from the~\textsc{Spindr} test set, here 5YEA. We then select three ligands at random and compare them to the reference compound based on Vina score, SA score, QED, logP, chemical similarity, and interaction similarity. Atom colors: C (cyan/gray), N (blue), O (red), S (yellow), F (ochre), Cl (green), H (white)}
    \label{fig:flowr_func_group_cond_5yea}
\end{figure}

Here we show additional results on the protein target with PDB ID 5YEA for different conditional modes using~\textsc{Flowr.multi}.

\FloatBarrier
\newpage
\renewcommand{\refname}{Supplementary References}

\end{document}